\documentclass[12pt,draftclsnofoot,onecolumn]{IEEEtran}
\usepackage[margin=1in]{geometry}
\usepackage{cite}
\usepackage{amsmath,amssymb,amsfonts}
\usepackage{algorithmic}
\usepackage{graphicx}
\usepackage{algorithm,algorithmic}
\usepackage{hyperref}
\usepackage{textcomp}
\usepackage{siunitx}
\usepackage{subcaption}
\def\eps{\varepsilon}
\def\beq{\begin{equation}}
\def\eeq{\end{equation}}

\def\di{\displaystyle}
\newcommand{\trn}{^{\scriptscriptstyle T}}
\newtheorem{theorem}{Theorem}

\begin{document}
\title{
Derivative Estimation from Coarse, Irregular, Noisy Samples: An MLE–Spline Approach} 
\author{Konstantin E. Avrachenkov \quad and\quad Leonid B. Freidovich
\thanks{Konstantin 
Avrachenkov is with INRIA Sophia Antipolis, France (e-mail: k.avrachenkov@inria.fr).}
\thanks{Leonid 
Freidovich is with the Robotics and Control Group, Department of Applied Physics and Electronics, Ume\r{a} University, Sweden (e-mail: leonid.freidovich@umu.se).}}

\maketitle

\begin{abstract}
We address numerical differentiation under coarse, non-uniform sampling and Gaussian noise. A maximum-likelihood estimator with $L_2$-norm constraint on a higher-order derivative is obtained, yielding spline-based solution. We introduce a non-standard parameterization of quadratic splines and develop recursive online algorithms. Two formulations—quadratic and zero-order—offer tradeoff between smoothness and computational speed. Simulations demonstrate superior performance over high-gain observers and super-twisting differentiators under coarse sampling and high noise, benefiting systems where higher sampling rates are impractical.
\end{abstract}
\section{Introduction}

Numerical differentiation is a fundamental task in control, signal processing, monitoring, and estimation. In many applications, one does not observe the derivative of a signal directly, but only noisy measurements of the signal itself, taken at discrete time instants. The difficulty is especially pronounced when the samples are not evenly spaced, when the sampling rate is relatively low compared with the signal dynamics, and when the measurement noise is significant. These features are common in practical systems, where sampling rates are constrained by sensing, communication, or computation limits rather than chosen solely for estimation accuracy.

Specifically, we consider the following problem. A continuous-time signal is observed only through noisy samples taken at known but possibly irregular time instants. The objective is to reconstruct its time derivative from these samples. We assume that the underlying signal possesses a prescribed degree of smoothness, expressed through an $L_2$ bound on a higher-order derivative. The central question is therefore how to combine the measurement model and the smoothness constraint into an estimator that is both theoretically justified and practically implementable, with a preference for real-time operation.

We study this problem in two settings. The first is the fixed-time interval setting, in which all samples on a given observation window are available and the derivative is reconstructed over the whole interval. This is the natural batch formulation and is the setting in which one can exploit all available information. The second is the recursive setting, motivated by online operation. In this case, samples arrive sequentially and the derivative estimate must be updated as new data become available. The recursive setting is particularly important for control and monitoring tasks, where decisions are taken in real time and recomputing a full batch solution after each new sample may be too costly.

The existing literature offers a wide spectrum of differentiators, each addressing in a different way the fundamental trade-offs between estimation delay, noise sensitivity, computational complexity, and modeling assumptions. Since this literature is vast, we restrict our discussion to the main classes of differentiators and cite only representative articles and surveys.

Causal Finite Impulse Response (FIR) differentiators, such as backward differences, causal  Savitzky-Golay filters \cite{gorry1990general,diop2000numerical,SavGol}, or causal algebraic differentiators \cite{Mboup2009numerical,Mboup2018frequency,Othmane2022survey} are frequently employed for their simplicity and lack of internal dynamics, yet they often introduce significant phase lag and struggle with noisy measurements, and non-uniform and irregular sampling. These drawbacks become acute when the sampling is coarse or significantly irregular. Noncausal or fixed-lag FIR variants, such as centered finite differences, centered Savitzky–Golay filters, or fixed-lag algebraic techniques \cite{savitzky1964smoothing,gorry1990general,SavGol} use future as well as past samples. These methods often achieve better accuracy and lower phase distortion than purely causal schemes. At the same time, they require either look-ahead or an explicit delay, are usually designed for uniform grids, and remain affected by boundary effects.
Causal Infinite Impulse Response (IIR) filters and low-pass differentiation schemes \cite{Alaoui2007} are computationally efficient and easy to deploy in embedded control loops, with tuning often expressed in terms of a bandwidth or time constant. Their weakness is that reducing phase lag without strongly amplifying noise is difficult, and their tuning depends on the sampling rate. This dependence is especially problematic when the grid is irregular.

Another important family is model-based state estimation, including Luenberger observers \cite{luenberger1979introduction}, high-gain observers \cite{gauthier1992simple, estefanidiari1992output, tornambe1993high,besancon2000remarks,dabroom2001,Vasiljevic2008,
Khalil2017HighGain%
}, and Kalman filtering methods \cite{kalman1960filter,belanger1998velocity}. When an accurate dynamical model is available, such approaches can provide excellent causal estimates and a principled treatment of noise. However, their performance relies on the quality of the assumed model; model mismatch may produce large bias or poor transients.

There is also an influential line of nonlinear robust differentiators, in particular sliding-mode and super-twisting schemes, pioneered by Levant \cite{Levant1998robust}, see also \cite{fridman2011,Barbot2020discrete,Moreno2023,acary2010implicit,Livne,levant2020robust, Carvajal2022contribution, Carvajal2022implicit,
seeber2024implicit, aldana2025optimal}. These methods are appealing because of their robustness properties, finite-time behavior, and relatively simple structure. Note that the sliding mode differentiator asymptotically achieves the fundamental limit on accuracy \cite{Kolmogorov1962}.  However, they typically require conservative bounds on higher-order derivatives and 
careful tuning against noise. They can also suffer from long transients. 

{Finally, one should mention offline or batch regularization-based methods, such as smoothing splines \cite{nemirovski1984signal,wahba1990spline}, Tikhonov regularization \cite{Cullum1971numerical},  basis selection methods \cite{meng2022smoothing} or total-variation regularization \cite{Chartrand2011numerical}.} These methods are often among the most accurate approaches in low signal-to-noise settings. Their main limitations are noncausality, computational cost, and the need to choose regularization parameters and basis representations carefully.
We also note that although one may estimate the signal itself by a Tikhonov- or spline-type smoothing procedure and then differentiate the reconstruction, this indirect route does not 
target the derivative as the primary estimation object. By contrast, as we shall see, a direct formulation in terms of the derivative exploits the stable integral relation between the derivative and the observations, imposes regularization directly on the quantity of interest, and naturally leads to economic parameterizations and recursive updates.

The aim of this paper is to combine the main advantages of these different families of methods while avoiding some of their most restrictive assumptions. Specifically, we would like our estimator to satisfy the following ``wish list'' of properties: it should have a solid theoretical foundation, admit both causal and noncausal forms, be amenable to recursive updates, cope with unbounded Gaussian noise, exhibit short transients, handle coarse and irregular sampling, and remain easy to tune.

Our first contribution is a rigorous formulation of derivative estimation under coarse, non-uniform sampling and additive Gaussian noise as a constrained variational maximum-likelihood problem over a Sobolev space. For this problem, we establish existence and uniqueness of the solution and show that the optimizer is a spline of an appropriate order. This gives the method a precise statistical and functional-analytic foundation, rather than presenting it only as a heuristic filtering rule. In particular, our Lagrangian reformulation provides a way to compute and to interpret the regularization parameter as a Lagrange multiplier. 
We note that in smoothing splines the regularization parameter is typically chosen in an ad hoc manner.
The second contribution is algorithmic: for quadratic and zero-order splines, we derive finite-dimensional convex quadratic optimization problems with solutions given by linear matrix equations. 
The third contribution introduces a non-standard parameterization of quadratic splines that reduces the number of parameters relative to classical spline constructions. 
The fourth contribution is the development of recursive updating formulas, enabling online implementation without recomputing the full solution from scratch whenever a new sample arrives. 
Finally, simulation results show that the proposed estimators perform competitively under coarse and noisy sampling and outperform discretized high-gain observers and super-twisting differentiators in many regimes.

The main advance over the state of the art is not only a new estimator, but a new combination of properties that were not available simultaneously. Compared with classical finite-difference and FIR/IIR schemes, the proposed method is explicitly designed for coarse and irregular sampling and is grounded in a regularized likelihood principle. Compared with observer-based differentiators, it does not require a detailed dynamical model. With respect to sliding-mode-based method, our solution avoids 
transients and is naturally matched to Gaussian measurement noise. Compared with standard batch spline or regularization methods, it comes with recursive implementations tailored to online estimation. The zero-order variant further offers a particularly simple alternative that avoids boundary conditions and appears attractive in online setup, while the quadratic variant provides greater smoothness.

\section{Non-uniform noisy sampling of a continuous-time 
signal\label{sec:theory}}


\subsection{Problem formulation\label{sec:problem}}
Let 
$
x(t),\ t \in [0,T],\ 0<T<\infty,
$
denote a continuous-time signal and
\beq\label{z(t)}
z(t) := \hbox{$\frac{d}{dt}$}x(t),\qquad t\in [0,T],
\eeq
denote its derivative.
The signal is sampled non-uniformly at known time instances 
\beq\label{t_k}
t_k \in [0,T],\quad k=1,\dots,{K}:\quad 0=t_1
<\dots<t_{K} = T.
\eeq
Furthermore, we receive noisy measurements of the signal
\beq\label{y(t_k)}
y(t_k) = x(t_k) + n_k, \quad n_k \sim N\bigl(0,\sigma^2\bigr),\quad k=1,\dots,{K},
\eeq
where $N\bigl(0,\sigma^2\bigr)$ denotes the Gaussian law with zero mean and variance $\sigma^2 \not= 0$,  
although
other types of noise can also be considered. 

Our goal is to reconstruct the derivative from the observed noisy samples.
Since numerical differentiation is an ill-posed problem, one needs to introduce regularization. Specifically,
we assume that the signal $x(t)$ is $(m+1)$-times differentiable  and its higher order derivative satisfies 
the following bound
\beq
\label{eq:boundL2} 
\frac1T\int_{0}^{T}\bigl(z^{(m)}(\tau)\bigr)^2\,d\tau \le L_{m+1}^2, 
\eeq
where the derivatives are understood in the weak (almost-everywhere) sense.
%
Following \cite{Kolmogorov1962}, in \cite{Levant1998robust} a uniform bound on a higher derivative was proposed as a regularization. Condition \eqref{eq:boundL2} is a relaxation.

\subsection{Maximum log-likelihood estimator as a variational problem\label{sec:likelihood}}
We denote the sampling intervals 
\beq\label{sampling_intervals}
h_k=t_{k+1}-t_{k} > 0,\quad k=1,\dots,{K-1}.
\eeq
Let us also denote
\beq
\label{z_k}
y_k=y(t_k), \qquad z_k=z(t_k), \qquad k=1,...,K.
\eeq

Our goal is to estimate the derivative $z(t)$, see (\ref{z(t)}),
$t \in [0,T]$, based on the sample times $t_k$ and the sample values $y_k$, for $k=1,\dots,K$.
We are interested either in the values at the sample moments, $z_k$, or at some other moments of time.

Using the Leibniz-Newton formula, i.e. integrating (\ref{z(t)}), we can write
$$
y_k=x(t_k)+n_k=x(t_1)+\int_{t_1}^{t_k} z(\tau)\,d\tau + n_k.
$$
The latter means that
\beq\label{cont_distr}
y_k \sim N\left(x(t_1)+\int_{t_1}^{t_k} z(\tau)\,d\tau, \sigma^2 \right),\quad k=1,\dots,{K}.
\eeq  

To find $x(t_1)$ and $z(t),$ $t \in [0,T],$ we propose to use the maximum log-likelihood estimator.
To this end, we first write the likelihood
\begin{align*}
f\bigl(y\,|\,z,x(t_1)\bigr) &\propto 
\exp\left(-\frac{\left( y_1 - x(t_1) \right)^2}{2\sigma^2}\right) \times \cdots 
\times \exp\left(-\frac{1}{2\sigma^2}\left( y_K - x(t_1) - \displaystyle\int_{t_1}^{t_K} z(\tau)\,d\tau \right)^2\right)
\end{align*}
and then take its logarithm with the minus sign
$$
-\log(f) \propto
\sum_{k=1}^{K} \left(y_{k}-x(t_1)-\int_{t_1}^{t_k} z(\tau)\,d\tau \right)^2.
$$
Thus, we need to solve the following variational problem
\beq
\label{eq:ContDiscML}
\min_{z\in H^m_{[0,T]},\,x(t_1)} \sum_{k=1}^{K} \left(y_{k}-x(t_1)-\int_{t_1}^{t_k} z(\tau)\,d\tau \right)^2,
\eeq
subject to constraint 
(\ref{eq:boundL2}), and
where 
$H^m_{[0,T]}$ is the $m$-order Sobolev space with the Euclidean norm. 
%
%
Let us next characterize the solution of the variational problem (\ref{eq:ContDiscML}).

\medskip

\medskip

\begin{theorem}\label{Theorem_1}\em
Suppose that $K > m$, $L_{m+1}>0$, and there is no exact interpolation that satisfies the constraint (\ref{eq:boundL2}). Then, the variational problem (\ref{eq:ContDiscML}) subject to constraint (\ref{eq:boundL2}) has a finite unique solution that can be obtained from the Lagrangian reformulation
\beq\label{eq:MinL_Opt}
\min_{x(t_1),\,z\in H^{m}_{[0,T]}}L\bigl(x(t_1),z(\cdot)\bigr), 
\eeq
where $[t_1,t_K]\equiv[0,T]$ and
\begin{align}
\label{eq:Lagr_Opt}
L\bigl(x_0,z(\cdot)\bigr) &= 
\sum_{k=1}^{K} \left(y_{k} - x_0 - \int_{t_1}^{t_k} z(\tau)\,d\tau \right)^2 
+ \lambda \left( \int_{0}^{T} \Bigl(z^{(m)}(\tau)\Bigr)^2\,d\tau 
- L_{m+1}^2\,T \right).
\end{align}
%
The Lagrange multiplier $\lambda_* > 0,$ which corresponds to the constraint satisfaction, can be found from the dual problem
\begin{align}
\label{eq:Lagr_Opt_dual}
\lambda_* = \arg\max_{\lambda \ge 0} \min_{x(t_1),\,z \in H^{m}_{[0,T]}} 
&\sum_{k=1}^{K} \!\left(y_{k} - x(t_1) - \int_{t_1}^{t_k}\! z(\tau)\,d\tau \right)^2 
 + \lambda \left( \int_{0}^{T} \bigl(z^{(m)}(\tau)\bigr)^2\,d\tau - L_{m+1}^2\,T \right).
\end{align}

\noindent
Moreover, the optimal solution for (\ref{eq:Lagr_Opt}) is a $2m$-order spline.
\medskip
\end{theorem}
For proof, see the 
Appendix.

\section{Optimization of spline parameters\label{sec:optimization}}

We proceed to formalize the analytical expression of the optimal derivative approximation, derived in Theorem~\ref{Theorem_1}, into a numerical procedure, treating $\lambda$ as a tuning parameter and not as a computed optimal value.  We consider the cases $m=0$ and $m=1$. Since the case $m=1$ is more involved, we present it first and in greater detail. The case of $m=0$ will follow with a straightforward adaptation.

\subsection{Quadratic splines}

Let us consider the case of $m=1$ in Theorem~\ref{Theorem_1} and the quadratic spline defined by functions $z_i(t)$ for $t \in [t_i, t_{i+1}]$ for $i=1,\dots, K-1$ as:
\beq\label{matching_quadratic_spline}
z_i(t) =   \frac{(t - t_i)^2}{h_i^2} z_{i+1} 
+ \frac{(t - t_i)(t - t_{i+1})}{h_i^2}  p_i 
+ \frac{(t - t_{i+1})^2}{h_i^2}  z_i, 
\eeq
which  satisfy $z_i(t_i)=z_i$ and $z_i(t_{i+1})=z_{i+1}$ so that $z_i(t_{i+1})=z_{i+1}(t_{i+1})$ for $i=1,\dots,(K-2)$. 
These polynomials are parameterized by $(2\,K-1)$ parameters $\{z_i\}_{i=1}^K$ and $\{p_i\}_{i=1}^{K-1}$. 
Geometrically, $-p_i/2$ is the middle Bernstein-B\'ezier control point of the quadratic segment \cite{prautzsch2002bezier}. So $p_i$ determines how strongly the quadratic curve is pulled away from the straight line joining $z_i$ and $z_{i+1}$.

Differentiating with respect to $t$, we obtain
\[
z_i'(t) = \frac2{h_i^2}(t-t_i)(z_{i+1}+z_{i}+p_{i}) -\frac{p_{i}+ 2\,z_{i}}{h_i}
\]
\[
\quad\hbox{and}\qquad z_i''(t) = \frac2{h_i^2}(z_{i+1}+z_{i}+p_{i}).
\]
To ensure continuity of the first derivative, we need to satisfy the following $(K-2)$ conditions:
\[
z_i'(t_{i+1})=\frac{2\,z_{i+1}+p_i}{h_i}
=-\frac{2\,z_{i+1}+p_{i+1}}{h_{i+1}}=z_{i+1}'(t_{i+1}),
\]
$i=1,\dots,K\!-\!2.$
Solving this expression for $z_{i+1}$, we obtain
\beq\label{ZPmap}
z_{i+1}=-\frac12\frac{p_{i+1}\,h_{i}+p_{i}\,h_{i+1}}{h_{i}+h_{i+1}}.
\eeq


If we impose the boundary conditions
$z_1''(t_{1}+)=0$ and $z_{K-1}''(t_{K}-)=0$, we obtain
\begin{align}
\label{zeroend2der}
z_1 &= \di - z_2 - p_1 
= \frac{\frac{h_1}{2}p_2 - \left(\frac{h_2}{2} + h_1\right)p_1}{h_1 + h_2}, \notag \\
z_K &= \di - z_{K-1} - p_{K-1} 
= \frac{\frac{h_{K-1}}{2}p_{K-2} - \!\left(\!\frac{h_{K-2}}{2}\! +\! h_{K-1}\!\right)p\!_{K-1}}{h_{K-2}+! +\! h_{K-1}}.
\end{align}

Equations (\ref{ZPmap}) and (\ref{zeroend2der})
result in the linear relation (\ref{PtoZmap}).
{\small
\beq\label{PtoZmap}
\left[\begin{array}{c} z_1 \cr z_2 \cr z_3 \cr \vdots \cr z_{K-1} \cr z_K \end{array}\right]=
\left[\begin{array}{cccccc} 
-\frac{2h_1+h_2}{2(h_1+h_2)} & \frac{h_1}{2(h_1+h_2)} & 0 &\cdots& 0 & 0\cr 
-\frac{h_2}{2(h_1+h_2)} & -\frac{h_1}{2(h_1+h_2)} & 0 & \cdots& 0 & 0\cr 
0& -\frac{h_3}{2(h_2+h_3)} & -\frac{h_2}{2(h_2+h_3)} & \cdots& 0 & 0 \cr 
\vdots \cr 
0&0&0& \cdots& -\frac{h_{K-1}}{2(h_{K-2}+h_{K-1})} & -\frac{h_{K-2}}{2(h_{K-2}+h_{K-1})} \cr 
0&0&0& \cdots& \frac{h_{K-1}}{2(h_{K-1}+h_{K-2})}& -\frac{2h_{K-1}+h_{K-2}}{2(h_{K-1}+h_{K-2})}
\end{array}\right]
\left[\begin{array}{c} p_1 \cr p_2 \cr p_3 \cr \vdots \cr p_{K-2} \cr p_{K-1}\end{array}\right]
\eeq
}

The objective in (\ref{eq:ContDiscML}) can be rewritten in the following form:
\begin{align}
\sum_{k=1}^{K} \left(y_{k} - x(t_1) - \int_{t_1}^{t_k} z(\tau)\,d\tau \right)^2 
&= \left(y_{1} - x(t_1)\right)^2 \label{for_reviewer}
+ \sum_{k=2}^{K} \left(y_{k} - x(t_1) 
- \sum_{i=1}^{k-1} \int_{t_i}^{t_{i+1}} z_i(\tau)\,d\tau \right)^2
.\end{align}
Let us substitute $z_i(t)$ from \eqref{matching_quadratic_spline} into each integral in \eqref{for_reviewer}:
\begin{align}
\int_{t_i}^{t_{i+1}} z_i(\tau) \, d\tau 
&= \frac{1}{h_i^2} \Bigg[
\int_{t_i}^{t_{i+1}} z_{i+1}(\tau - t_i)^2 \, d\tau \notag \\
&\hskip-20mm+\! \int_{t_i}^{t_{i+1}} p_i (\tau - t_i)(\tau - t_{i+1}) \, d\tau 
 +\! \int_{t_i}^{t_{i+1}} z_i(\tau - t_{i+1})^2 \, d\tau \Bigg] 
 = \frac{1}{h_i^2} \left( \frac{z_{i+1} h_i^3}{3} - \frac{p_i h_i^3}{6} + \frac{z_i h_i^3}{3} \right)
\end{align}
and, as a result, we obtain
the following optimization problem:
\begin{align}
\label{quadratic_zp}
\min_{\{z_i\}_{i=1}^K,\,\{p_i\}_{i=1}^{K-1},\,x(t_1)} \Bigg\{ 
&\left(y_1 - x(t_1)\right)^2 
+ \sum_{k=2}^{K} \left( y_k - x(t_1) 
- \sum_{i=1}^{k-1} \left( \frac{z_{i+1} h_i}{3} 
- \frac{p_i h_i}{6} + \frac{z_i h_i}{3} \right) \right)^2 
\Bigg\},
\end{align}
subject to (\ref{ZPmap}), (\ref{zeroend2der}), and a condition implied by (\ref{eq:boundL2}).


There are several ways how the optimization problem above can be converted to a least-square-like problem. 
%
%
 Substituting (\ref{PtoZmap}) into (\ref{quadratic_zp}), we obtain an equivalent problem
\beq\label{quadratic}
\min_{\{p_i\}_{i=1}^{K-1},\,x(t_1)} 
\left\{\Bigl(C_K\,Z_K-Y_K\Bigr)\trn\,\Bigl(C_K\,Z_K-Y_K\Bigr)\right\},
\eeq
subject to a condition implied by (\ref{eq:boundL2}),
with the corresponding matrix $C_K\in{\mathbb R}^{K\times K}$
and
\begin{align}
Z_K &= \Bigl[\!\begin{array}{cccc} x(t_1), & p_1, & \cdots, & p_{K-1} \end{array}\!\Bigr]\trn, \label{ZK}\\
Y_K &= \Bigl[\!\begin{array}{cccc} y_1, & y_2, & \cdots, & y_K \end{array}\!\Bigr]\trn.\label{YK}
\end{align}
Next, using (\ref{PtoZmap}), we can compute the singular matrix $Q_K$: 
\begin{align}
\label{QK}
\int_{0}^{T} \bigl(z'(\tau)\bigr)^2\,d\tau 
&= \sum_{i=1}^{K-1} \int_{t_i}^{t_{i+1}} 
\Bigl( \frac{2}{h_i^2}(t - t_i)(z_{i+1} + z_i + p_i)  
- \frac{p_i + 2z_i}{h_i} \Bigr)^2 d\tau
= Z_K\trn\, Q_K\, Z_K.
\end{align}
Using Theorem~\ref{Theorem_1} with $m=1$
, we obtain the 
unconstrained optimization problem:
\begin{align}
\label{quadratic_relaxed}
\min_{Z_K} \Bigl\{ 
&\frac{1}{K}\bigl(C_K Z_K - Y_K\bigr)\trn \bigl(C_K Z_K - Y_K\bigr) 
+ \lambda \Bigl( \frac{1}{t_K - t_1} Z_K\trn Q_K Z_K- (L_2)^2 \Bigr)\Bigr\}.
\end{align}
Since $-\lambda L_2^2$ is a constant offset with respect to $Z_K$, fixing $\lambda > 0$ as our single tuning parameter and rescaling $\lambda\to\frac{t_K-t_1}{K}\,{\lambda}$ allow us to rewrite (\ref{quadratic_relaxed}) as
$\min_{Z_K} \Bigl\{ 
\bigl(C_K Z_K - Y_K\bigr)\trn \bigl(C_K Z_K - Y_K\bigr) 
+\lambda Z_K\trn Q_K Z_K 
\Bigr\}
$ 
the solution of which coincides with the solution of the linear system
\beq\label{lin_sys}
A_K\,Z_K=\bigl(C_K\bigr)\trn Y_K
,\quad 
A_K=\bigl(C_K\bigr)\trn C_K+\lambda\,Q_K.
\eeq

\subsection{Zero-order splines}	

Let us now consider the simpler case of $m=0$ of Theorem~\ref{Theorem_1} and the piecewise constant function defined by $z_i(t)$ for $t \in [t_i, t_{i+1})$ for $i=1,\dots, K-1$ as:
\begin{equation} \label{matching_zero_spline}
    z_i(t) = z_{i},\qquad t_{i+1}=t_i+h_i,
\end{equation}
and $z_{K-1}(t_{K})=z_{K-1}$.

Proceeding in the conceptually the same way as above, one obtains (\ref{lin_sys}) with (\ref{YK}),
\begin{align*}
Z_K&=\Bigl[\!\begin{array}{cccc}x(t_1), & z_1,& \cdots,& z_{K-1}\end{array}\!\Bigr]\trn,\\
C_K &= \begin{bmatrix}
1 & 0 & 0 & \cdots & 0 \\
1 & h_1 & 0 & \cdots & 0 \\
1 & h_1 & h_2 & \cdots & 0 \\
\vdots & \vdots & \vdots & \ddots & \vdots \\
1 & h_1 & h_2 & \cdots & h_{K-1}
\end{bmatrix}, \notag \\
Q_K &=\mathop{\rm diag}\bigl\{0,~h_1,~\cdots,~h_{K-1} \bigr\}
.
\end{align*}

\section{Recursive implementation}\label{sec:rec}

As we mentioned earlier, most control tasks are executed over an increasing time horizon and require online estimation of the derivative. Thus, we propose a recursive method to solve the system (\ref{lin_sys}) with newly arriving samples.
Similarly to the previous section, we begin with the harder case of $m=1$ and, after that, we summarize the result for $m=0$. 


\subsection{Quadratic splines}
Suppose that we have solved the problem (\ref{lin_sys}) for $m=1$ and a particular sufficiently large $K$. More specifically, suppose that we have computed $A_K^{-1}$ for
\[
A_K= C_K \trn C_K+\lambda\,Q_K\in {\mathbb R}^{K\times K}
,\quad
\hat{Z}_K=A_K^{-1} C_K \trn Y_K\in {\mathbb R}^{K},
\]
where $\hat{Z}_K\!=\!\Bigl[\hat x(t_1),~\hat p_1,\dots,\hat p_{K-1}\Bigr]\trn$ is defined as $Z_K$ in (\ref{ZK}).

Let us derive the recursive update expressions for  $Z_{K+1}=A_{K+1}^{-1} C_{K+1} \trn Y_{K+1}$, using the newly acquired measurement $y_{K+1}$, avoiding explicit inversion of the matrix $A_{K+1}$. Towards this goal, we find the updates $\Delta Z$ and $p_K$ defining the solution of the new optimization problem
\begin{align}
\label{ZK+1}
Z_{K+1} 
&
=\!\Bigl[  x(t_1), p_1,\cdots, p_{K} \Bigr]\trn
\!\!= \!\Bigl[(\hat Z_K + \Delta Z)\trn, p_K \Bigr]\trn\\
&\hskip-10mm= \Bigl[\hat x(t_1) + \Delta x(t_1),~ \hat p_1 + \Delta p_1,~ \dots,~ 
\hat p_{K-1} + \Delta p_{K-1},~ p_K \Bigr]\trn\!. \notag 
\end{align}

Using symbolic computations, we obtain
\begin{align}
\label{CQK+1}
C_{K+1} &= 
\begin{bmatrix}
C_K & 0_{K \times 1} \\
0_{1 \times K} & 0
\end{bmatrix}
+
\begin{bmatrix}
0_{(K-1) \times (K+1)} \\
\Delta C
\end{bmatrix}, \\
Q_{K+1} &= 
\begin{bmatrix}
Q_K & 0_{K \times 1} \\
0_{1 \times K} & 0
\end{bmatrix}
+
\begin{bmatrix}
0_{(K-2) \times (K-2)} & 0_{(K-2) \times 3} \\
0_{3 \times (K-2)} & \Delta Q
\end{bmatrix}, \notag
\end{align}
where $0$ denotes a matrix of zeros of appropriate dimension and for $K\ge 5$, we have $\Delta C \in {\mathbb R}^{2\times (K+1)}$ 
below 
\begin{figure*}[!t]
\centering
{\small Definition of several entries of correction matrix \(\Delta C\) for the extended system, see \cite{MLEDiff}:}
{\tiny
\begin{equation*}
\Delta C\!=\!
\begin{bmatrix}
0 & \cdots & 0 &
-\dfrac{h_{K-2}^2}{6(h_{K-3}+h_{K-2})} &
\dfrac{h_{K-2}^2(h_{K-3}\!+\!2h_{K-2}\!+\!h_{K-1})}{6(h_{K-3}\!+\!h_{K-2})(h_{K-2}\!+\!h_{K-1})} &
-\dfrac{h_{K-2}^2}{6(h_{K-2}\!+\!h_{K-1})} \\
1 &
-\dfrac{3h_1^2\! +\! 3h_1 h_2\! +\! h_2^2}{6(h_1 \!+\! h_2)} &
-\dfrac{2h_1 h_2 \!+\! h_1 h_3 \!+\! h_2 h_3 \!+\! h_2^2}{6(h_1 \!+\! h_2)} &
\cdots &
\cdots &
-\dfrac{h_{K-2}^2\! +\! 3h_{K-2}h_{K-1}\! +\! 3h_{K-1}^2}{6(h_{K-2}\!+\!h_{K-1})}
\end{bmatrix}\!.
\end{equation*}}
\vspace{-1em}
\end{figure*}
(see GitHub \cite{MLEDiff} for the detailed expressions of the entries of the second row of $\Delta C$), while $\Delta Q=\Delta Q\trn\in {\mathbb R}^{3\times 3}$ is defined by
\begin{align*}
\Delta Q_{1,1} &= -\frac{2h_{K-2}}{3(h_{K-3}+h_{K-2})^2}, \notag \\
\Delta Q_{1,2} &= \frac{h_{K-2}(h_{K-3}+5h_{K-2}+4h_{K-1})}{6(h_{K-3}+h_{K-2})^2(h_{K-2}+h_{K-1})}, \notag \\
\Delta Q_{1,3} &= -\frac{h_{K-2}}{6(h_{K-3}+h_{K-2})(h_{K-2}+h_{K-1})}, \notag \\
\Delta Q_{2,2}&=
\frac{h_{K-2}^2 + 4h_{K-2}h_{K-1} + h_{K-3}h_{K-2} + h_{K-1}^2 + 3h_{K-3}h_{K-1}}{3(h_{K-3}+h_{K-2})(h_{K-2}+h_{K-1})^2} 
- \frac{h_{K-3}+3h_{K-2}}{3(h_{K-3}+h_{K-2})^2}, \notag \\ \qquad 
\Delta Q_{2,3}&= 
 -\frac{2h_{K-3}h_{K-2} + 6h_{K-3}h_{K-1} + 5h_{K-2}h_{K-1} + h_{K-2}^2}{6(h_{K-3}+h_{K-2})(h_{K-2}+h_{K-1})^2}, \notag \\
\Delta Q_{3,3} &= \frac{h_{K-2} + 3h_{K-1}}{3(h_{K-2}+h_{K-1})^2}.
\end{align*}

The update from $K$ to $K{+}1$ is obtained by using the structure induced by appending one new interval. The expressions \eqref{CQK+1} allow $A_{K+1}\in{\mathbb R}^{(K+1)\times(K+1)}$ to be written as the sum of a zero-padded extension of $A_{K}\in{\mathbb R}^{K\times K}$ and an additional matrix $\Delta A$. The matrix $A_{K+1}$ is partitioned into four blocks by separating the last row and the last column. The inverse of the large square block, denoted by $A_s$, is computed using a sequence of low-rank updates based on the structure of $\Delta A$. The block matrix inversion formula is then applied to compute $A_{K+1}^{-1}$. In parallel, the same block decomposition yields update equations for the initial value $x(t_1)$, the spline parameters $p_1,\dots,p_{K-1}$ for the previous intervals, and the new spline parameter $p_K$, associated with the added interval. The resulting update relations are summarized below.

\begin{theorem}\em
Given $K\ge 5$, $y_K$, $y_{K+1}$, $A_K^{-1}$, and $\hat Z_K$; $Z_{K+1}$, defined by (\ref{ZK+1}), can be computed using the formulae:
\begin{align}
\label{DeltaZ}
\Delta Z &= \left(A_s^{-1} - \frac{A_s^{-1} U_s U_s\trn A_s^{-1}}{U_s\trn A_s^{-1} U_s - a_s} \right)
\begin{bmatrix} b_1 \!\!&\!\! \dots\!\! &\!\! b_K \end{bmatrix}\trn 
- \frac{b_{K+1}}{a_s} U_s, \notag \\
p_K &= \frac{b_{K+1} - U_s\trn \Delta Z}{a_s},
\end{align}
where $\begin{bmatrix} b_1  & \dots & b_K & b_{K+1}\end{bmatrix}\trn= \Delta C\trn \begin{bmatrix} y_K \\ y_{K+1} \end{bmatrix} - \Delta A\begin{bmatrix} \hat Z_K \\ 0 \end{bmatrix}$, $\Delta A$ is defined by the right-hand side of (\ref{DeltaA}) below, while
$A_s^{-1}$ can be obtained using 
three rank-two updates with the Sherman-Morrison-Woodbury formula for 
\begin{align}
\label{As}
A_s = A_K 
&+ \left\{ \Delta C\trn I_{2\times 2}(\Delta C + C_s) \right\}_{\star}
+ \left\{ C_s\trn I_{2\times 2} \Delta C \right\}_{\star} \notag \\
&+ \begin{bmatrix} 0_{(K-2)\times 2} \\ I_{2\times 2} \end{bmatrix}
\left( \lambda\, \Delta Q_s \right)
\begin{bmatrix} 0_{(K-2)\times 2} \\ I_{2\times 2} \end{bmatrix}\trn,
\end{align}
where the operation $\Bigl\{\cdots\Bigr\}_{\star}=\begin{bmatrix} I_{K\times K} \cr 0_{1\times K}\end{bmatrix}\trn\Bigl\{\cdots\Bigr\}\begin{bmatrix} I_{K\times K} \cr 0_{1\times K}\end{bmatrix}$ extracts the upper left corner block of size $K\times K$ and ${\Delta Q}_s=\begin{bmatrix} I_{2\times 2} \cr 0_{1\times 2}\end{bmatrix}\trn\Delta Q \begin{bmatrix} I_{2\times 2} \cr 0_{1\times 2}\end{bmatrix}$ is the upper left $2\times 2$ block of $\Delta Q$, while $C_s\!=\!\begin{bmatrix} \bigl(C_K\bigr)_{1,K} &  \cdots & \bigl(C_K\bigr)_{K,K} & 0 \\ 0 &  \cdots & 0 & 0 \end{bmatrix}$ is zero-padded extension of the last row of $C_K$ with $\Delta C$ and $\Delta Q$ are from (\ref{CQK+1}).
Moreover, $A_{K+1}^{-1}$ can be computed using $A_s^{-1}$, see (\ref{AK+1inv}) below.
\medskip
\end{theorem}
{\bf Proof:}
The updated matrix, that would be used to directly compute ${Z}_{K+1}=A_{K+1}^{-1}C_{K+1}\trn Y_{K+1}$ or solving the corresponding linear systems of higher order, is $A_{K+1} = C_{K+1}\trn C_{K+1} + \lambda Q_{K+1}$.
To avoid recomputing the inverse of $A_{K+1}$ or solving the corresponding linear system, let us substitute (\ref{CQK+1}) into the latter expression and exploit the structure of $\Delta C$. As a result, the correction matrix can be computed as follows:
\begin{align}
\label{DeltaA}
\Delta A &= 
\begin{bmatrix} A_s& U_s \\ U_s\trn & a_s \end{bmatrix}-
\begin{bmatrix}
A_K & 0_{K \times 1} \\
0_{1 \times K} & 0
\end{bmatrix} \notag \\
&= (\Delta C)\trn (\Delta C) + (\Delta C)\trn C_s + C_s\trn (\Delta C) 
+ \lambda \begin{bmatrix}
0_{(K-2)\times(K-2)} & 0_{(K-2)\times3} \\
0_{3\times(K-2)} & \Delta Q
\end{bmatrix},
\end{align}
with\vskip-5mm
\[ 
C_s\trn(\Delta C)=\begin{bmatrix}
C_K & {0}_{K\times 1} \\
{0}_{1\times K} & 0
\end{bmatrix}\trn
\begin{bmatrix}
  0_{(K-1)\times (K+1)}\cr
\Delta C
\end{bmatrix},
\]
where $A_{K+1}\in \mathbb{R}^{(K+1)\times (K+1)}$ is divided into four blocs: 
$A_s=A_s\trn \in \mathbb{R}^{K\times K}$, $U_s \in \mathbb{R}^{K}$ and $a_s \in \mathbb{R}$ and $0_{(\cdot)}$ denote zero matrices of appropriate dimensions.

Since $A_{K+1}$ is defined by (\ref{DeltaA}) as a sum of two (or four) singular matrices, we cannot directly use the Sherman-Morrison-Woodbury formula:
$$\left( A_0 + U_0 C_0 V_0\trn \right)^{-1} = A_0^{-1} - A_0^{-1} U_0 \left( C_0^{-1} + V_0\trn A_0^{-1} U_0 \right)^{-1} V_0\trn A_0^{-1},$$ see e.g. \cite[pp.\,65–66]{golub2013matrix}.
However, with the above division into blocks of $A_{K+1}$, one can use the (somewhat related) block matrix inversion formula \cite[pp.\,28–29]{golub2013matrix}
\begin{align}
\label{AK+1inv}
A_{K+1}^{-1} &=
\begin{bmatrix}
A_s^{-1} + d_s A_s^{-1} U_s U_s\trn A_s^{-1} & -d_s A_s^{-1} U_s \\
-d_s U_s\trn A_s^{-1} & d_s
\end{bmatrix}, 
\qquad
d_s 
= \frac{1}{a_s - U_s\trn A_s^{-1} U_s},
\end{align}
if $A_s^{-1}$ was computed. Next, let us rewrite the equation $A_{K+1}Z_{K+1}=C_{K+1}\trn Y_{K+1}$ as
\begin{align}
&\left(
\begin{bmatrix}
A_K & 0 \\
0 & 0
\end{bmatrix}
+ \Delta A
\right)
\left(
\begin{bmatrix}
\hat Z_K \\
0
\end{bmatrix}
+
\begin{bmatrix}
\Delta Z \\
p_K
\end{bmatrix}
\right) 
= C_{K+1}\trn Y_{K+1} =
\begin{bmatrix}
C_K\trn Y_K \\
0
\end{bmatrix}
+
\begin{bmatrix}
0 \\
(\Delta C)\trn 
\begin{bmatrix}
y_K & y_{K+1}
\end{bmatrix}\trn
\end{bmatrix},
\end{align}
and substitute $C_K\trn Y_K=A_K\hat Z_K$. 
If we solve the latter equation for $p_K$ first, and then substitute the solution into the first $K$ equations for $\Delta Z$, and solve them, we obtain
the updates (\ref{DeltaZ}). 


It is left to notice that $A_s$  is defined by the left upper $K\times K$ block of (\ref{DeltaA}) that can be written as (\ref{As}).
%
\hfill $\Box$\medskip

\subsection{Zero-order splines}
Suppose now we have solved the problem (\ref{lin_sys}) for $m=0$ and a particular 
$K$, i.e. we have computed $A_K^{-1}$ for
\[
A_K= C_K \trn C_K+\lambda\,Q_K\in {\mathbb R}^{K\times K},\quad 
\hat{Z}_K=A_K^{-1} C_K \trn Y_K\in {\mathbb R}^{K},
\]
where $\hat{Z}_K=\Bigl[\hat x(t_1),~\hat z_1,~\dots,~ \hat z_{K-1}\Bigr]\trn$ denotes the values obtained earlier, before $y_{K+1}$ was available.

It is straightforward to show that
\[
C_{K+1}=\begin{bmatrix}C_K ~~ 0_{K\times 1}\cr H_K\trn\end{bmatrix},\qquad 
Y_{K+1}=\begin{bmatrix}Y_K \cr y_{K+1}\end{bmatrix},\]\[
Q_{K+1}=\begin{bmatrix}Q_K & 0_{K\times1}\cr 0_{1\times K} & h_K\end{bmatrix},\quad
H_K=\begin{bmatrix}1 & h_1 & \dots & h_K\end{bmatrix}\trn ,
\]\[
A_{K+1}=\begin{bmatrix}A_K & 0_{K\times 1}\cr 0_{1\times K} & 0\end{bmatrix}\!+\!\left(\!H_K\,H_K\trn+\!\begin{bmatrix}0_{K\times K} & 0_{K\times 1}\cr 0_{1\times K} & \lambda\, h_K\end{bmatrix}\right), 
\]
and, therefore, 
\[
A_{K+1}^{-1}\!=\!A_s^{-1}-\frac{A_s^{-1}H_K H_K\trn A_s^{-1}}{1+H_K\trn A_s^{-1}H_K}, ~ A_s^{-1}\!=\!\begin{bmatrix}A_K^{-1} & 0_{K\times 1}\cr 0_{1\times K} & \frac{1}{\lambda\, h_K}\end{bmatrix},
\]
\[
Z_{K+1}=A_{K+1}^{-1}\,B_{K+1},\quad 
B_{K+1}=\begin{bmatrix}B_{K} \\ 0\end{bmatrix}+H_K\,y_{K+1}.
\]

\subsection{Sliding-window variation}

The recursive implementations proposed above require storing generally dense inverse matrices $A_K^{-1}$, whose dimensions grow with the number of measurements $K$. To bound memory and computational requirements, 
we suggest a sliding-window variation retaining only 
the latest $M$ measurements.
Upon arrival of a new measurement, the oldest one is removed
by downdating $A_K^{-1}$ and shifting the origin of the
integral representation. For quadratic splines ($m=1$),
removing the oldest observation and the regularization term
over the first interval leads to a rank-two downdate,
followed by a rank-one correction enforcing the boundary
condition $z''(t_2+)=0$. For zero-order splines ($m=0$),
no boundary correction is needed, and the downdate can be
obtained using a Schur complement and a rank-one correction.
Combining these operations with the recursive updates above yields an algorithm that reproduces the solution of the finite-window optimization problem, using only the latest $M$ measurements, with $O(M^2)$ memory and computational complexity per update.


\section{Numerical results\label{sec:numerical}}

Let us compare the performance of our two algorithms with the two simplest Euler-integration implementations of Levant's differentiator, also known as the super-twisting differentiator, which asymptotically achieves the best possible accuracy, see, e.g., \cite{Livne}, and a basic linear HGO, equivalent to a linear differential filter defined by a bandwidth parameter, followed by saturation, see, e.g., \cite{Khalil2017HighGain
}. 

We implemented the algorithms in MATLAB; the code is available in~\cite{MLEDiff}. As test data, we take $x(t)=t-1+\frac12\left(\frac{\sin(2\pi t)}{2\pi}+\frac{\sin(3.1\pi t)}{3.1\pi}\right)$ on $t \in [0,T]$ with $T=1.95$. We generate sampling intervals  (\ref{sampling_intervals}) independently as $h_k \sim \mathop{\rm Uniform}([0.5h,1.5h])$ and noise in (\ref{y(t_k)}) according to $n_k \sim N\bigl(0,\sigma^2\bigr)$, a zero-mean Gaussian distribution with standard deviation $\sigma$. We experiment with various combinations of sampling rate and noise levels, given by $h$ and $\sigma$, respectively.

We consider two scenarios: In the first scenario, we need to estimate the derivative over the entire time interval $[0,T]$ with fixed $T=1.95$; and in the second ``online'' scenario, given all the available measurements up to the current moment $T$, we need to estimate the derivative value only at $T$.
In the second scenario, we vary the value of $T$ from 0 to 1.95.
 
\begin{figure*}[ht]
\centering
\begin{subfigure}{0.95\linewidth}
    \includegraphics[width=\linewidth]{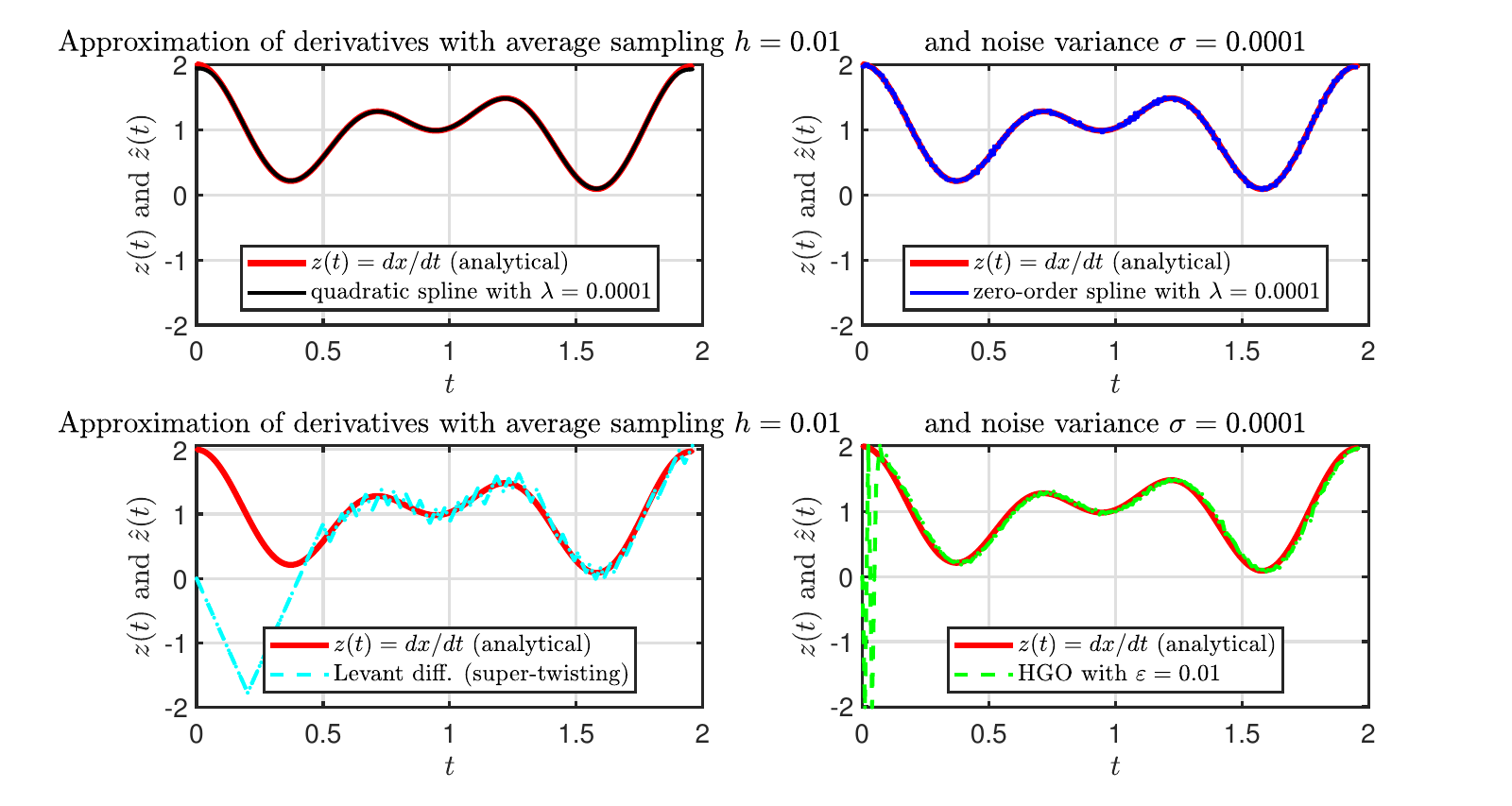}\vskip-2mm
    \caption{Lower average sampling rate (100 Hz)}\label{h0.01}
\end{subfigure}
\hfill
\begin{subfigure}{0.95\linewidth}
    \includegraphics[width=\linewidth]{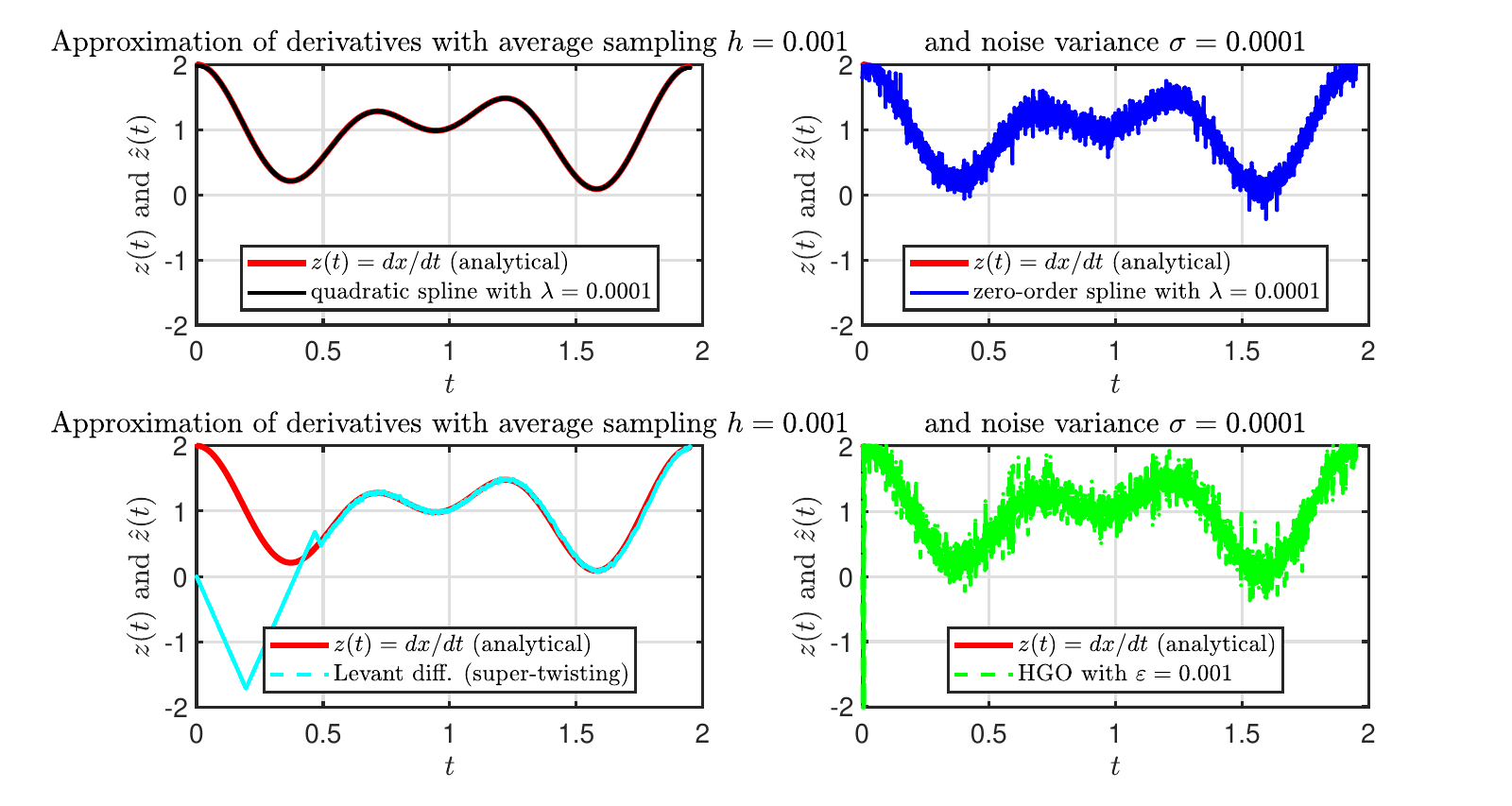}\vskip-2mm
    \caption{Higher average sampling rate (1000 Hz)}\label{h0.001}
\end{subfigure}
\caption{Comparison of performance under different average sampling rates.}
\label{fig:sampling}
\end{figure*}

Under lower sampling rate\footnote{In industrial control systems, the selection of sampling intervals is closely related to the time constants of the regulated processes. For instance, in industrial robotics, where joint dynamics exhibit time constants of 10 to 100ms, sampling intervals of 0.001s to 0.01s are typical~\cite{zheng2025robotics}. In chemical plants, where time constants range from several seconds to minutes, sampling intervals are typically between 0.1s and 1s~\cite{kenett2023process}.
}
at $100$\,Hz (Fig.~\ref{fig:sampling}\,(a)), both MLE-Spline methods show superior performance with an advantage of quadratic splines, which, however, require much higher computational and memory resources, even with a recursive implementation. Results under finer sampling rate at $1000$\,Hz (Fig.~\ref{fig:sampling}\,(b)), show that the simpler‑to‑implement Levant's algorithm has an advantage when transient behavior is not critical. 
  
\begin{figure*}[ht]
\centering
\begin{subfigure}{0.95\linewidth}
    \includegraphics[width=\linewidth]{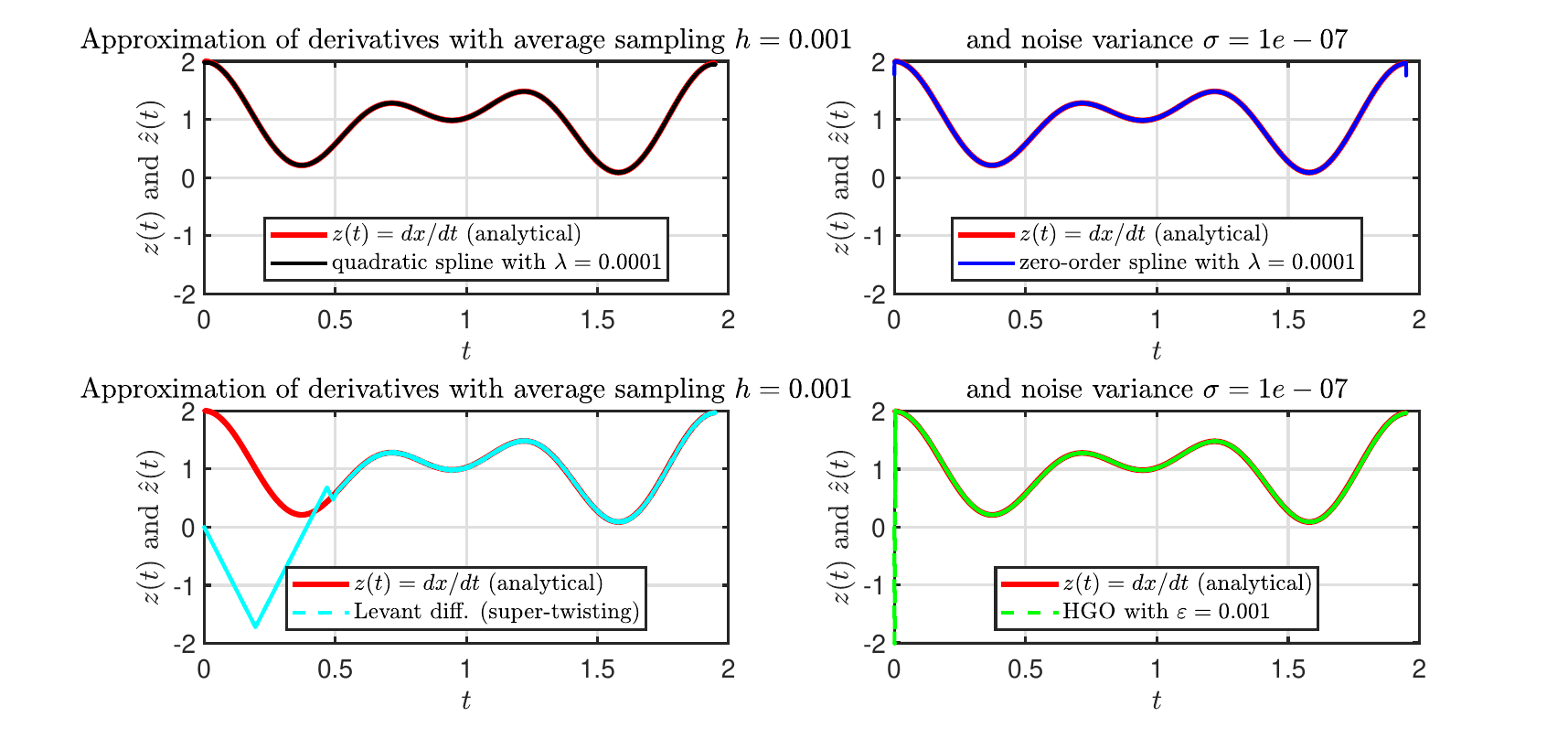}\vskip-2mm
    \caption{Insignificant noise}\label{h0.001lownoise}
\end{subfigure}
\hfill
\begin{subfigure}{0.95\linewidth}
    \includegraphics[width=\linewidth]{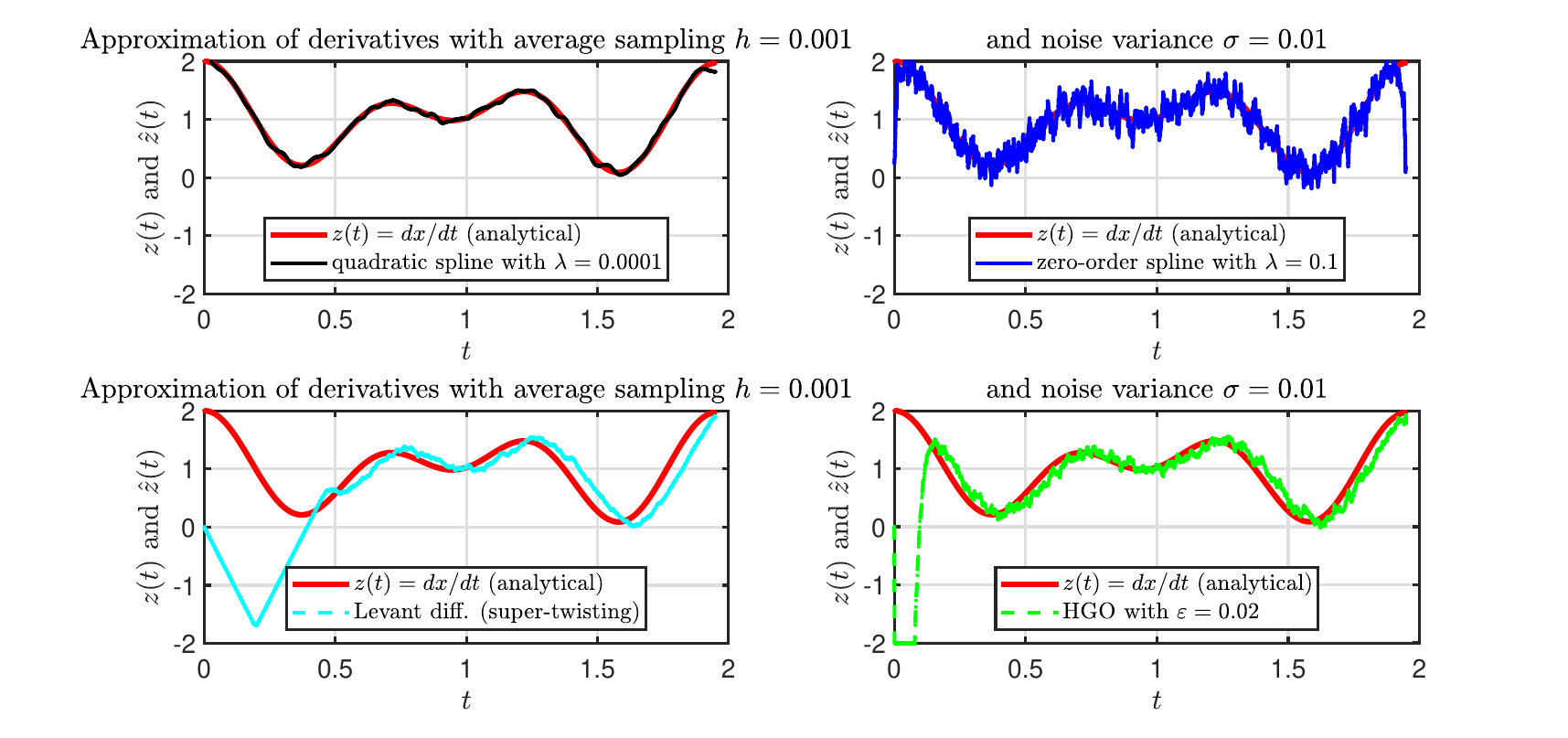}\vskip-2mm
    \caption{Significant noise}\label{h0.001highnoise}
\end{subfigure}
\caption{Comparison of performance under different levels of noise.}
\label{fig:noise}
\end{figure*}

\begin{figure}[ht]
\centering
\includegraphics[width=\linewidth]{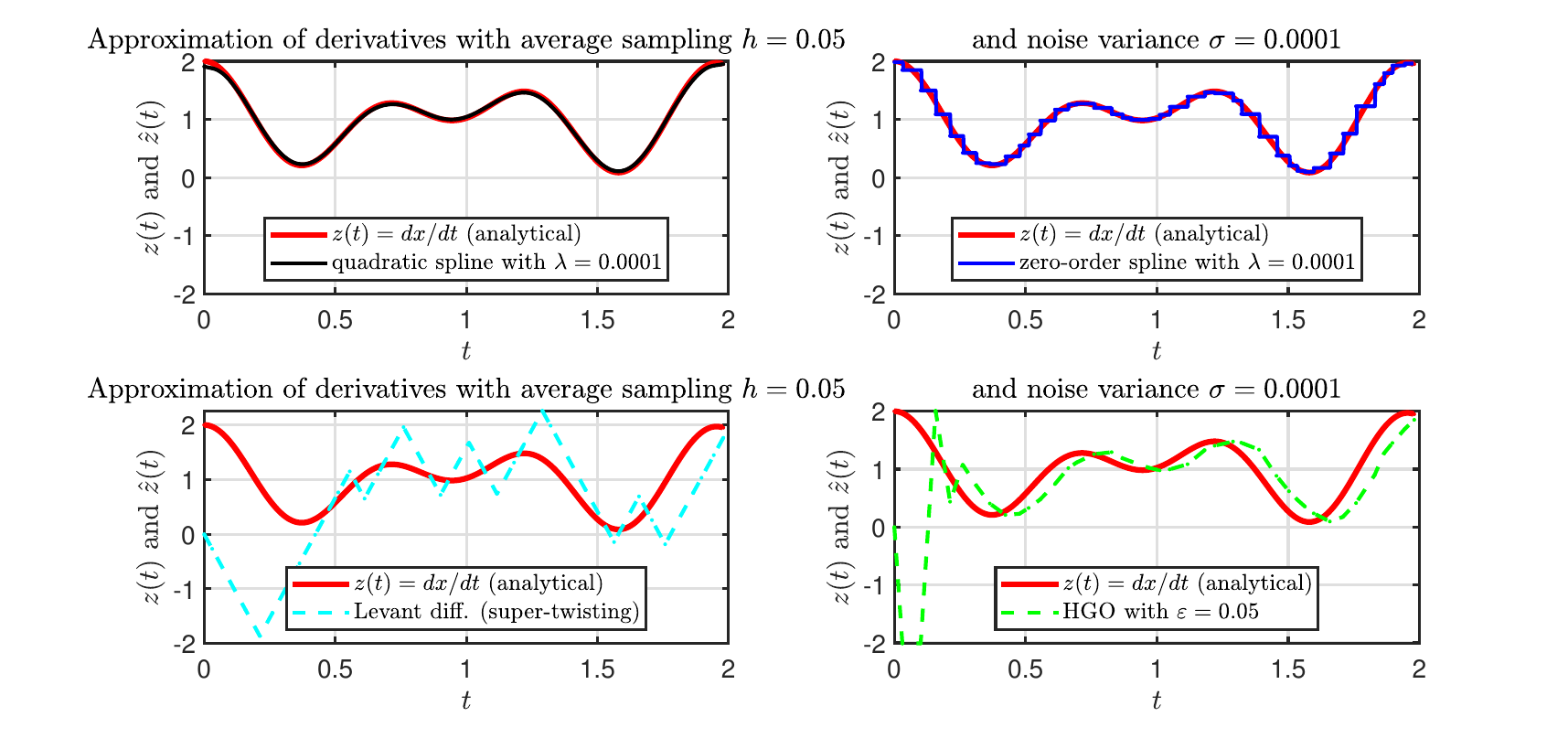}
\vskip-2mm
\caption{Performance under coarse non-uniform sampling.}\label{h0.05}
\end{figure}

In the following, we observe the performance of the algorithms under relatively fine sampling for control system applications. When the level of noise is extremely low, see Fig.
~\ref{fig:noise}\,(a), all three methods give very good estimates with poorer end-point performance for our two methods and worse transients for the other two. Note that, under very noisy measurements, see Fig.
~\ref{fig:noise}\,(b), zero-order splines perform poorly, while the other three methods behave similarly with a better transient when the quadratic splines are used.
 Finally, under coarse non‑uniform sampling (Fig.\,\ref{h0.05}), the MLE‑Spline methods achieve lower Root Mean Squared Errors (RMSEs)
.

Table~\ref{tab:results} reports the RMSEs in the full interval estimation scenario, computed over the final two-thirds of the time interval, corresponding to the steady-state regime after the initial transient phase.

\begin{table}[ht]
\caption{
RMSE after the transients in offline 
estimation 
}\vskip-1mm
\label{tab:results}
\centering
\resizebox{0.7\linewidth}{!}{%
\begin{tabular}{l
                S[table-format=1.4,detect-weight=true,mode=text]
                S[table-format=1.4,detect-weight=true,mode=text]
                S[table-format=1.4,detect-weight=true,mode=text]
                S[table-format=1.4,detect-weight=true,mode=text]}
\hline
\textbf{Parameters ($\mathbf{h}$,\,$\mathbf{\sigma}$,\,Fig.~\#)} & $\mathbf{m\!=\!0}$ & $\mathbf{m\!=\!1}$ & \textbf{Levant} & \textbf{HGO} \\
\hline
$0.0002,~10^{-7}$ & 0.0067 & \textbf{~\underline{0.0012}} & \textbf{~0.0018} & 0.0074\\
$0.001,~10^{-7}$,~Fig.~2a  & \textbf{~0.0041} & \textbf{~\underline{0.0031}} & 0.0085 & 0.0058\\
$0.001,~10^{-4}$,~Fig.~1b & 0.1157 & \textbf{~\underline{0.0032}} & \textbf{~0.0260} & 0.1582 \\
$0.001,~10^{-2}$,~ Fig.~2b & 0.1931  & $\textbf{~\underline{0.0413}}$ & 0.2276 & \textbf{~0.1641} \\ 
$0.01,~10^{-4}$,~ Fig.~1a & \textbf{~0.0259} & \textbf{~\underline{0.0101}} & 0.0907 & 0.0597 \\
$0.05,~10^{-4}$,~ Fig.~3 & \textbf{~0.1025} & \textbf{~\underline{0.0226}} & 0.3663 & 0.2705 \\
$0.075,~10^{-3}$  & \textbf{~0.1517} & \textbf{~\underline{0.0452}} & 0.7918 & 0.5533 \\
\hline
\end{tabular}
}
\vspace{10mm}

\parbox{\textwidth}{\normalsize{\em Note:} {\bf Boldface} indicates the two smallest RMSEs in each experiment (the best is \underline{underlined}). The super-twisting method uses standard parameters from \cite{Levant1998robust}; the optimal spline parameter was tuned by trial and error for one experiment and kept fixed ($\lambda=10^{-4}$) for the others (except for $0$-order spline under significant noise); and the HGO parameter ($\varepsilon>0$) was tuned by trial and error in each experiment (the best choice depends on sampling \cite{dabroom2001} and noise \cite{Vasiljevic2008}).}
\end{table}

Table~\ref{tab:online_results} reports the online scenario; RMSE is computed only at the rightmost point as the time horizon $T$ grows from 0 to 1.95s.

\begin{table*}[ht]
\centering

\begin{minipage}[t]{0.65\textwidth}
\centering
\vspace{0pt}

\captionof{table}{%
RMSE after the transients for online estimation
}
\label{tab:online_results}

\vskip-1mm

\resizebox{\linewidth}{!}{%
\begin{tabular}{
    l
    S[table-format=2.4,detect-weight=true,mode=text]
    S[table-format=2.4,detect-weight=true,mode=text]
    S[table-format=2.4,detect-weight=true,mode=text]
    S[table-format=2.4,detect-weight=true,mode=text]
}
\hline
\textbf{Parameters ($\mathbf{h}$,\,$\mathbf{\sigma}$)}
& $\mathbf{m\!=\!0}$
& $\mathbf{m\!=\!1}$
& {\textbf{Levant}}
& {\textbf{HGO}} \\
\hline

$0.0002,~10^{-7}$
& 0.3102
& 0.0507
& {\textbf{\underline{0.0018}}}
& {\textbf{0.0074}} \\

$0.001,~10^{-7}$
& 0.1055
& 0.0979
& {\textbf{0.0087}}
& {\textbf{\underline{0.0058}}} \\

$0.001,~10^{-4}$
& 0.1619
& {\textbf{0.0981}}
& {\textbf{\underline{0.0261}}}
& 0.1695 \\

$0.001,~10^{-2}$
& 1.0373
& {\textbf{\underline{0.1269}}}
& 0.2403
& {\textbf{0.1661}} \\

$0.01,~10^{-4}$
& {\textbf{\underline{0.0306}}}
& 0.1594
& 0.0772
& {\textbf{0.0600}} \\

$0.05,~10^{-4}$
& {\textbf{\underline{0.1058}}}
& {\textbf{0.1838}}
& 0.4848
& 0.2705 \\

$0.075,~10^{-3}$
& {\textbf{\underline{0.1721}}}
& {\textbf{0.1851}}
& 0.5370
& 0.5262 \\

\hline
\end{tabular}%
}

\end{minipage}
\bigskip\bigskip

{\begin{minipage}[t]{0.65\textwidth}
\centering
\vspace{20pt}
\captionof{table}{%
{RMSE for fixed-step and uniform noise}
}
\label{tab:adaptive_differences}

 \vskip-1mm

 \resizebox{\linewidth}{!}{%
 \begin{tabular}{
     l
     S[table-format=2.4,detect-weight=true,mode=text]
     S[table-format=2.4,detect-weight=true,mode=text]
     S[table-format=2.4,detect-weight=true,mode=text]|
     S[table-format=2.4,detect-weight=true,mode=text]
 }
 \hline
 \textbf{Parameters ($\mathbf{h}$, $\sigma$)}
 & $\mathbf{m\!=\!0}$
 & $\mathbf{m\!=\!1}$
 & {\cite{Seeber2023}}
 & {\textbf{offline}}\\
 \hline










%
%



%
$0.0002,~10^{-7}$
& 0.2877
& \textbf{0.0673}
& \textbf{\underline{0.0009}}
& 0.0022 \\

$0.001,~10^{-7}$
& \textbf{0.0899}
& 0.1000
& \textbf{\underline{0.0020}}
& 0.0036 \\

$0.001,~10^{-4}$
& 0.1459
& \textbf{0.1000}
& \textbf{\underline{0.0243}}
& 0.0037 \\

$0.001,~10^{-2}$
& 1.1108
& \textbf{\underline{0.1251}}
& \textbf{0.2420}
& 0.0280 \\

$0.01,~10^{-4}$
& \textbf{\underline{0.0251}}
& 0.1667
& \textbf{\underline{0.0251}}
& 0.0073 \\

$0.05,~10^{-4}$
& \textbf{0.1025}
& 0.2033
& \textbf{\underline{0.1020}}
& 0.0212 \\

$0.075,~10^{-3}$
& \textbf{\underline{0.1531}}
& 0.1803
& \textbf{0.1560}
& 0.0403 \\


\hline
\end{tabular}%
}

\end{minipage}
}
\end{table*}


{For equally spaced observations ($h_k=h=\mathrm{const}$) and bounded noise, \cite{Seeber2023} 
proposed a sample-based optimal robust differentiator based on finite differences 
with an adaptive denominator, under a uniform bound on the second derivative, which is the base differentiator of \cite{aldana2025optimal}. 
An attractive feature of this differentiator is its almost negligible transient 
behavior, in contrast to HGOs and sliding-mode differentiators.
Although the constraint in \cite{Seeber2023} differs from the time-averaged constraint considered in 
our approach, a numerical comparison is useful for assessing the practical performance 
of the two methods. We therefore compare the differentiator of \cite{Seeber2023} with our estimators in Table~\ref{tab:adaptive_differences} (``offline'' corresponds to the full interval and $m=1$). We consider the uniform noise, as in \cite{Seeber2023}, from the interval $\left[-\sqrt{3}\,\sigma,\sqrt{3}\,\sigma\right]$. By choosing the uniform noise, we satisfy
the boundedness assumption of \cite{Seeber2023} and test our methods against a non-Gaussian noise.
}

One can see a drastic difference between the two scenarios that can be explained by the fact that the optimal spline has a significantly increased estimation error close to the endpoints of the interval. 

It is important to remind the readers that the above-compared methods are developed under different assumptions. We also would like to report, without demonstrating results, that our methods appear to be not sensitive to the value of the single tuning parameter
$\lambda$, which, however, we recommend to tune for particular applications. The most sensitive, as well known, to parameters is the considered HGO while there are many results improving the super-twisting differentiator and its discrete-time implementations, see, e.g., \cite{acary2010implicit,Livne,Carvajal2022implicit,seeber2024implicit}. However, to the best of our knowledge, they consider bounded noise and assume fine sampling. The transient behavior of such algorithms can be speed up by simply increasing the parameter that must exceed the upper bound on the second derivative but such an increase would result in bigger chattering and higher sensitivity to the level of noise. 
Furthermore, we have also tried numerically B-splines \cite{prautzsch2002bezier}, their performance near interval boundaries was
unsatisfactory, likely due to error propagation.

Obviously, claiming the best performance based on numerical examples is not possible. At the same time, we believe that our method is demonstrated to be worth trying for practical cases, especially, under coarse, 
non-uniformly sampled noisy measurements.


\section{Conclusion and discussion\label{sec:conclusion}}

The paper has presented a rigorous framework for numerical differentiation under coarse, non-uniform sampling and additive Gaussian noise. We formulated the problem as a constrained variational optimization and proved the existence and uniqueness of a spline solution. We introduced a novel, economical parametrization of quadratic splines and developed a recursive online implementation. We also demonstrated the effectiveness of simpler zero-order splines.

Our method requires only one tuning parameter, and its performance is relatively insensitive to its choice. We found that boundary conditions strongly affect estimation quality near the end points, making zero-order splines, which require no boundary conditions, particularly attractive for online estimation.

Simulations demonstrated strong performance under coarse sampling and high noise levels,
particularly compared with discretized high-gain observers and super-twisting differentiators. While no method can claim universal superiority, our approach offers a compelling alternative for control applications constrained by hardware or sampling limitations.
\clearpage

\section*{Appendix: Proof of Theorem~\ref{Theorem_1}}

Let us first present the proof for the case $m\ge 1$. The optimization is considered in the space ${\mathbb R} \times H^{m}_{[0,T]}$,
where $H^{m}_{[0,T]}$ is the Sobolev space, with the
norm
\beq\label{norm}
\left\|\bigl(x_0,z(\cdot)\bigr)\right\|^2 = \bigl(x_0\bigr)^2 + \int_{0}^{T}\Bigl(\bigl(z(\tau)\bigr)^2 + \bigl(z^{(m)}(\tau)\bigr)^2 \Bigr)\, d\tau.
\eeq
This is a Hilbert space, which is reflexive. We use continuous
representatives for the elements of $H^{m}_{[0,T]}$ (for more background
on Sobolev spaces see, e.g., \cite{Brezis}).

The convexity of the main objective and constraint function as well as Slater's
condition imply the validity of the Lagrangian reformulation (see 
Chapter~8 of \cite{Luenberger}). The non-interpolation assumption, together with strong duality, implies that the Lagrange multiplier is positive.

Let us show that the Lagrangian in (\ref{eq:Lagr_Opt}) is in fact strictly convex.
Using the identity
$$
\bigl(\alpha a + (1-\alpha) b\bigr)^2 = \alpha a^2 + (1-\alpha) b^2 -\alpha(1-\alpha) (a-b)^2
$$
for $a\neq b$ and $0<\alpha<1$ for each term in (\ref{eq:Lagr_Opt})
and the linearity of integration, we need to show that if
$$
\sum_{k=1}^{K} \!\left(\!-\Delta x(t_1)\!-\!\int_{t_1}^{t_k}\!\! \Delta z(\tau)\,d\tau \!\right)^2 
+ \lambda_*\! \int_{0}^{T}\!\!\bigl(\Delta z^{(m)}(\tau)\bigr)^2\,d\tau = 0,
$$
then $\Delta x(t_1)=0$ and $\Delta z(t)=0$, $t \in [0,T]$. First, we note that
each 
term
must be zero. From
\quad$
\int_{0}^{T}(\Delta z^{(m)}\bigl(\tau)\bigr)^2\,d\tau = 0
$\quad
we conclude that $\Delta z(\tau)$ can only be a polynomial of order $m-1$.
The term with $k=1$ immediately implies that $\Delta x(t_1) = 0$. The function
\quad$
a(t) = \int_{t_1}^t \Delta z(\tau)\, d\tau
$\quad
is a polynomial of degree $m$. Since $\bigl(a(t_k)\bigr)^2=\left(\displaystyle\int_{t_1}^{t_k}\Delta z(\tau)\,d\tau\right)^2=0$, it has $K>m$ roots on the interval $[0,T]$.
Hence, this polynomial must be zero.

Let us now show that if $K > m$, the Lagrangian is coercive,
i.e., if $\|(x(t_1),z)\| \to \infty$ then $L\bigl(x(t_1),z\bigr) \to +\infty$. We can prove coercivity
by contradiction. Suppose $L\bigl(x(t_1),z\bigr) \le C$ for some finite bound $C$.
It is immediate to conclude that \quad$|x(t_1)|$\quad and
\quad$
\int_{0}^{T}(z^{(m)}(\tau))^2\,d\tau
\quad$
are bounded. 
By integration by parts we make the following transformation
$$
\int_{t_1}^{t_k} z(\tau) \, d\tau =
(t_k-t_1)z(t_1)+...+\frac{1}{m!}(t_k-t_1)^mz^{(m-1)}(t_1) 
+ \frac{1}{m!} \int_{0}^{T} (t_k-\tau)^m_+ z^{(m)}(\tau) \, d\tau,
$$
where $(a)_+=\max\{a,0\}$. By Cauchy-Schwartz inequality, we can bound
$$
$$

Thus, we have
$$
\sum_{k=2}^{K} \!\left(\!\tilde{y}_{k}\!-\! (t_k\!-\!t_1)z(t_1)\!-\!...\!-\!\frac{1}{m!}(t_k\!-\!t_1)^mz^{(m-1)}(t_1) \!\right)^2 \!\le C,
$$
where $\tilde{y}_{k}$ are bounded. We can rewrite the above inequality in the matrix form
\quad$
(\tilde{y}-W\tilde{z})\trn (\tilde{y}-W\tilde{z}) \le C,
$\quad
with 
\quad$
\tilde{z} = \Bigl[z(t_1),\dots,z^{(m-1)}(t_1)\Bigr]\trn
$\quad
and
$$
W=\left[\begin{array}{ccc}
(t_2-t_1) & \cdots & \frac{1}{m!}(t_2-t_1)^m\\
\vdots & & \vdots\\
(t_K-t_1) & \cdots & \frac{1}{m!}(t_K-t_1)^m
\end{array}\right].
$$
Since 
$W$ is of the Vandermonde-type, it has full column rank and consequently non-zero
smallest singular value if $K-1 \ge m$. Thus, the vector $\tilde{z}$ is bounded if $K > m$. 
Now, using the Taylor series expansion and the triangle inequality, we can deduce the boundedness 
of the $L^2$-norm of $z$, namely,
$$
\|z(\tau)\|_{L^2} \le |z(t_1)|(t_K-t_1)^{1/2} + |z'(t_1)| \frac{(t_K-t_1)^{3/2}}{\sqrt{3}} +...
+\frac{1}{(m-1)!}\left\|\int_{t_1}^t (t-\tau)^{(m-1)} z^{(m)}(\tau) \, d\tau\right\|_{L^2}, 
$$
where all $|z^{(i)}(t_1)|$, $i=0,...,m-1$ on the right hand side are bounded by the preceding considerations. 
Consider now the reminder term. We first apply Cauchy-Schwartz inequality
$$
\left|\int_{t_1}^t (t-\tau)^{(m-1)} z^{(m)}(\tau) \, d\tau\right| \le \left( \int_{t_1}^t |t-\tau|^{2(m-1)} d\tau \right)^{1/2} \left( \int_{t_1}^t (z^{(m)}(\tau))^2 d\tau \right)^{1/2}
$$
$$
= \left( \frac{(t-t_1)^{2m-1}}{2m-1} \right)^{1/2} \|z^{(m)}\|_{L^2_{[t_1,t]}}
\le \left( \frac{(t-t_1)^{2m-1}}{2m-1} \right)^{1/2} \|z^{(m)}\|_{L^2_{[0,T]}}.
$$

Since the integral of a squared bounded function is bounded, we conclude that
$\|z(\tau)\|_{L^2}$ is bounded as well.
Thus, $\|(x(t_1),z)\|$ is bounded which contradicts the initial assumption that $\|(x(t_1),z)\| \to \infty$.

Next, we show that the objective is continuous.
Using Cauchy-Schwartz inequality for $H^{m}_{[0,T]}$ with the natural inner product defined by (\ref{norm}), we can write
$$
\left|x(t_1) + \int_{t_1}^{t_k} z(\tau) \, d\tau \right| \le \sqrt{1+(t_k-t_1)} \sqrt{(x(t_1))^2 
+ \int_{t_1}^{t_k} (z(\tau))^2 \, d\tau}
$$
$$
\le \sqrt{1+(t_k-t_1)} \sqrt{(x(t_1))^2 + \int_{t_1}^{t_k} (z(\tau))^2 + (z^{(m)}(\tau))^2 \, d\tau}.
$$
The penalty term with the derivative is continuous as a semi-norm.
Hence, the Lagrangian is continuous as a combination of continuous functions.
Now we apply Proposition~1.2 from Chapter~2 of \cite{EkelandTemam}, which says that the optimization problem
has a unique finite solution if the functional is strictly convex, lower-semi-continuous and coercive,
and defined on a reflexive space.

Once we justified that the problem is well-defined and has a solution, we can apply the technique
of Lagrange multiplier to characterize the form of the solution. Towards this goal, it is convenient
to introduce auxiliary variables
\begin{equation}
\label{eq:epsk_constr}
\eps_k = y_k - x(t_1) - \int_{t_1}^{t_k} z(\tau) \, d\tau, \quad k=1,...,K.
\end{equation}
Let us also use the integration by parts to make the following transformation
$$
\int_{t_1}^{t_k} z(\tau) \, d\tau =
(t_k-t_1)z(t_1)+...+\frac{1}{m!}(t_k-t_1)^mz^{(m-1)}(t_1) 
+ \frac{1}{m!} \int_{0}^{T} (t_k-\tau)^m_+ z^{(m)}(\tau) \, d\tau
$$
with $(a)_+=\max\{a,0\}$, and substitute the above expression into (\ref{eq:epsk_constr}).
Then, the problem (\ref{eq:MinL_Opt}) can be rewritten in the following equivalent form
\begin{equation}
\label{eq:Lagr_Opt_equiv}
\min_{x(t_1),
\,z\in H^{m}_{[0,T]}} 
\sum_{k=1}^{K} \eps_k^2 
+ \lambda_* \left(\int_{0}^{T}(z^{(m)}(\tau))^2\,d\tau - L_{m+1}^2\,T\right),
\end{equation}
subject to constraints (\ref{eq:epsk_constr}). 

We can handle constraints (\ref{eq:epsk_constr})
by including them in the Lagrangian. Specifically, now the Lagrangian takes the form
\begin{align}
L(z, x(t_1), \tilde{z}, \varepsilon_1, \dots, \varepsilon_K)=
\sum_{k=1}^{K} \varepsilon_k^2 
+ \lambda_* \left( \int_{0}^{T} \left( z^{(m)}(\tau) \right)^2 \, d\tau \right) - \lambda_* L_{m+1}^2 T \notag \\
&\hskip-110mm + \sum_{k=1}^{K} \lambda_k \bigl( y_k - x(t_1) - p(\tilde{z}) \bigr) 
- \sum_{k=1}^{K} \lambda_k \left( \frac{1}{m!} \int_{0}^{T} (t_k - \tau)^m_+ z^{(m)}(\tau) \, d\tau + \varepsilon_k \right),
\label{eq:newLagr}
\end{align}
where $\tilde z$ is defined 
above as a vector of values of $z(t)$ and its derivatives up to order $(m-1)$ at $t=t_1=0$
and $p(\tilde z)=(t_k-t_1)z(t_1)+\dots+\frac{1}{m!}(t_k-t_1)^mz^{(m-1)}(t_1)$.


Computing the first variation of the Lagrangian we obtain
\begin{align*}
\delta L 
&= 2 \sum_{k=1}^{K} \varepsilon_k \, \delta \varepsilon_k 
+ 2 \lambda_* \int_{0}^{T} z^{(m)}(\tau) \, \delta z^{(m)}(\tau) \, d\tau \\
&\quad + \sum_{k=1}^{K} \lambda_k \left( -\delta x(t_1) - \nabla p(\tilde{z}) \, \delta \tilde{z} \right) 
- \sum_{k=1}^{K} \lambda_k \left( \frac{1}{m!} \int_{0}^{T} (t_k - \tau)^m_+ \delta z^{(m)}(\tau) \, d\tau 
+ \delta \varepsilon_k \right)
\end{align*}
and the first-order optimality from arbitrariness of the variations of independent variables:
\begin{align*}
\delta \varepsilon_k, \quad k = 1, \dots, K 
&: \quad 2 \varepsilon_k = \lambda_k, \quad k = 1, \dots, K, \\
\delta x(t_1) 
&: \quad \sum_{k=1}^{K} \lambda_k = 0, \\
\delta z^{(j)}(t_1), \quad j = 0, \dots, m-1 
&: \quad \sum_{k=1}^{K} \lambda_k \frac{(t_k - t_1)^{j+1}}{(j+1)!} = 0
\end{align*}
\begin{align*}
\forall \delta z^{(m)} \in C_{[0,T]}\!: \quad
\int_{0}^{T}\!\! \Bigg( 2 \lambda_* z^{(m)}(\tau)
&- \frac{1}{m!} \sum_{k=1}^{K} \lambda_k (t_k - \tau)^m_+ \Bigg) 
\delta z^{(m)}(\tau) \, d\tau = 0
\end{align*}

From the latter condition and the fundamental lemma of the calculus of variations (see, e.g., \cite[Lemma 1.1.1]{jost1998calculus}),
we conclude that
$$
z^{(m)}(\tau) = \frac{1}{2\lambda_*m!} \sum_{k=1}^{K} \lambda_k (t_k-\tau)^m_+,
$$
and hence
$$
z(\tau) = \sum_{i=0}^{m-1}c_{i}(\tau-t_1)^i + \frac{(-1)^m}{2\lambda_*(2m)!} \sum_{k=1}^{K} \lambda_k (t_k-\tau)^{2m}_+ .
$$
Therefore, the function $z(\tau)$ is a $2m$-order spline. The new constants $c_j$, $j=0,\dots, m-1$, are linked by relations 
\[
z^{(j)}(t_1)=j!\,c_j + \frac{(-1)^j}{2\lambda_*(2m-j)!} \sum_{k=1}^{K} \lambda_k (t_k-t_1)^{2m-j}
\]
Thus, we have $2K\!+\!2m\!+\!1$ variables and $K\!+\!2m\!+\!1$ constraints. We can consider $\lambda_k$, $k=1,\dots, K,$ as free variables that can be determined by a dual optimization problem. 


Let us now turn to the case $m=0$.
In this case, $H^0_{[0,T]}=L^2_{[0,T]}$. 
Denote $B := L_1^2T > 0$.
For any $x_0\in\mathbb R$ and $z\in L^2_{[0,T]}$,
define
$$
w_k=x_0+\int_0^{t_k}z(t)\,dt,
\quad k=1,\ldots,K.
$$
Then, the objective in (\ref{eq:ContDiscML}) becomes
$\sum_{k=1}^K(y_k-w_k)^2$. Moreover, by the
Cauchy--Schwarz inequality,
$$
\int_0^T z(t)^2\,dt
=\sum_{i=1}^{K-1}\int_{t_i}^{t_{i+1}}z(t)^2\,dt
\geq \sum_{i=1}^{K-1}
\frac{1}{h_i}
\left(\int_{t_i}^{t_{i+1}}z(t)\,dt\right)^2
=\sum_{i=1}^{K-1}
\frac{(w_{i+1}-w_i)^2}{h_i}.
$$
Equality holds if and only if $z$ is constant
almost everywhere on each interval
$(t_i,t_{i+1})$.

Conversely, for any $w=(w_1,\ldots,w_K)$,
the choice
$$
x_0=w_1,\quad
z(t)=\frac{w_{i+1}-w_i}{h_i},
\quad t\in(t_i,t_{i+1}),
$$
satisfies the defining relations for $w_k$ and
attains equality in the preceding inequality.
Thus, the original variational problem is
equivalent, in optimal value, to
\begin{eqnarray}
\begin{aligned}
\min_{w\in\mathbb R^K}\quad&
F(w):=\sum_{k=1}^K(y_k-w_k)^2,\\
\text{subject to}\quad&
G(w):=\sum_{i=1}^{K-1}
\frac{(w_{i+1}-w_i)^2}{h_i}\leq B.
\end{aligned}
\label{eq:finite_opt_m0}
\end{eqnarray}
The feasible set is nonempty, closed, and convex,
while $F$ is continuous, coercive, and strictly
convex. Hence (\ref{eq:finite_opt_m0}) admits a unique minimizer
$w^*$.

The non-interpolation assumption is equivalent
to $G(y)>B$, since exact interpolation requires
$w=y$. We claim that the constraint is active
at $w^*$. Indeed, if $G(w^*)<B$, then $w^*$
would also be an unconstrained local minimizer
of $F$. Since the unique unconstrained
minimizer is $w=y$, this would contradict
$G(y)>B$. Consequently,
\begin{equation}
G(w^*)=B.
\label{eq:tight_constr_m0}
\end{equation}

We now establish uniqueness for the original
variational problem. Any feasible pair $(x_0,z)$
induces a feasible vector $w$ in (\ref{eq:finite_opt_m0}), with the
same objective value. Therefore, any minimizer
of the original problem must induce $w=w^*$.
Furthermore, the Cauchy–Schwarz inequality established above, together with the original energy constraint and the equality (\ref{eq:tight_constr_m0}), yields
$$
B \geq \int_0^T z(t)^2\, dt \geq G(w^*)=B.
$$
Equality in the Cauchy--Schwarz inequality
therefore implies
\[
z(t)=\frac{w_{i+1}^*-w_i^*}{h_i}
\quad\text{for a.e. }t\in(t_i,t_{i+1}).
\]
Together with $x_0=w_1^*$, this proves the
existence and uniqueness of the original
minimizer (with $z$ understood up to equality
almost everywhere). In particular, the
minimizer is a zero-order spline.

It remains to justify the Lagrangian formulation.
Since $B>0$, Slater's condition holds: any
constant vector $w$ satisfies $G(w)=0<B$.
Therefore, strong duality holds and the
Karush--Kuhn--Tucker conditions guarantee
the existence of an optimal multiplier
$\lambda^*\geq0$ such that $w^*$ minimizes
\[
F(w)+\lambda^*(G(w)-B).
\]
If $\lambda^*=0$, the optimality condition
would give $w^*=y$, contradicting $G(y)>B$.
Hence $\lambda^*>0$.

Finally, for every $\lambda\geq0$, replacing
$z$ by the piecewise constant function having
the same integrals over the sampling intervals
leaves the residual sum of squares unchanged
and does not increase the penalty.
Consequently,
$$
\begin{aligned}
&\inf_{x_0\in\mathbb R,\;z\in L^2_{[0,T]}}
\left\{
\sum_{k=1}^K
\left(y_k-x_0-\int_0^{t_k}z(t)\,dt\right)^2
+\lambda\left(\int_0^Tz(t)^2\,dt-B\right)
\right\}\\
&\qquad=
\min_{w\in\mathbb R^K}
\left\{F(w)+\lambda(G(w)-B)\right\}.
\end{aligned}
$$
Thus, the finite-dimensional dual problem
coincides with (\ref{eq:Lagr_Opt_dual}). 


\bibliography{NumDiff_s}

@book{Khalil2017HighGain,
  author    = {Khalil, Hassan K.},
  title     = {High-Gain Observers in Nonlinear Feedback Control},
  publisher = {Society for Industrial and Applied Mathematics},
  address   = {Philadelphia, PA},
  year      = {2017},
}

@article{gorry1990general,
  title={General least-squares smoothing and differentiation by the convolution ({S}avitzky-{G}olay) method},
  author={Gorry, Peter A},
  journal={Anal. Chem.},
  volume={62},
  number={6},
  pages={570--573},
  year={1990}
}

@article{meng2022smoothing,
  title={Smoothing splines approximation using {H}ilbert curve basis selection},
  author={Meng, Cheng and Yu, Jun and Chen, Yongkai and Zhong, Wenxuan and Ma, Ping},
  journal={J. Comput. Graph. Stat.},
  volume={31},
  number={3},
  pages={802--812},
  year={2022}
}

@book{prautzsch2002bezier,
  title={B{\'e}zier and {B}-spline {T}echniques},
  author={Prautzsch, Hartmut and Boehm, Wolfgang and Paluszny, Marco},
  year={2002},
  publisher={Springer}
}

@article{levant2020robust,
  title={Robust exact filtering differentiators},
  author={Levant, Arie and Livne, Miki},
  journal={Eur. J. Control},
  volume={55},
  pages={33--44},
  year={2020},
  publisher={Elsevier}
}

@book{wahba1990spline,
  title={Spline Models for Observational Data},
  author={Wahba, Grace},
  year={1990},
  publisher={SIAM}
}

@article{acary2010implicit,
  title={Implicit {E}uler numerical scheme and chattering-free implementation of sliding mode systems},
  author={Acary, Vincent and Brogliato, Bernard},
  journal={Syst. Control Lett.},
  volume={59},
  number={5},
  pages={284--293},
  year={2010},
  publisher={Elsevier}
}

@article{Alaoui2007,
  author={Al-Alaoui, Mohamad Adnan},
  journal={IEEE Trans. Signal Process.}, 
  title={Linear Phase Low-Pass {IIR} Digital Differentiators}, 
  year={2007},
  volume={55},
  number={2},
  pages={697-706}}

@article{savitzky1964smoothing,
  title={Smoothing and differentiation of data by simplified least squares procedures.},
  author={Savitzky, Abraham and Golay, Marcel JE},
  journal={Anal. Chem.},
  volume={36},
  number={8},
  pages={1627--1639},
  year={1964}
}

@article{zheng2025robotics,
  author    = {Bin Zheng and Lu Huang},
  title     = {Real-time monitoring and prediction applications of industrial robots using machine learning},
  journal   = {Int. J. Intell. Robot. Appl.},
  volume    = {9}, 
  pages     = {1915--1933},
  year      = {2025}
}

@incollection{kenett2023process,
  author    = {Ron S. Kenett and Shelemyahu Zacks and Peter Gedeck},
  title     = {Basic Tools and Principles of Process Control},
  booktitle = {Statistics for Industry, Technology, and Engineering},
  pages     = {11--57},
  year      = {2023},
  publisher = {Springer},
}

@book{jost1998calculus,
  title={Calculus of Variations},
  author={Jost, J{\"u}rgen and Li-Jost, Xianqing},
  year={1998},
  publisher={Cambridge University Press}
}

@book{luenberger1979introduction,
  author    = {Luenberger, David G.},
  title     = {Introduction to Dynamic Systems: Theory, Models, and Applications},
  publisher = {John  and Sons},
  address   = {New York},
  year      = {1979}
}

@book{golub2013matrix,
  author    = {Gene H. Golub and Charles F. Van Loan},
  title     = {Matrix {C}omputations},
  edition   = {4},
  publisher = {Johns Hopkins Univ. Press},
  year      = {2013},
  address   = {Baltimore},
}

@article{kalman1960filter,
  author    = {R. E. Kalman},
  title     = {A New Approach to Linear Filtering and Prediction Problems},
  journal = {J. Basic Eng.},
  year      = {1960},
  volume    = {82},
  number    = {1},
  pages     = {35--45},
  doi       = {10.1115/1.3662552}
}

@article{gauthier1992simple,
  author    = {Jean-Paul Gauthier and Hassen Hammouri and S. Othman},
  title     = {A Simple Observer for Nonlinear Systems: Applications to Bioreactors},
  journal = {IEEE Trans. Autom. Control},
  volume    = {37},
  number    = {6},
  pages     = {875--880},
  year      = {1992}
}

@article{estefanidiari1992output,
  author    = {F. Esfandiari and Hassan K. Khalil},
  title     = {Output feedback stabilization of fully linearizable systems},
  journal = {Int. J. Control },
  volume    = {56},
  number    = {5},
  pages     = {1007--1037},
  year      = {1992},
  doi       = {10.1080/00207179208934353}
}

@article{tornambe1993high,
  author    = {Antonio Tornambè},
  title     = {Output feedback stabilization of a class of nonminimum-phase nonlinear systems},
  journal = {Syst. Control Lett.},
  volume    = {29},
  pages     = {193--204},
  year      = {1992},
}

@article{besancon2000remarks,
  author    = {Gildas Besançon},
  title     = {High-gain observation with disturbance attenuation and application to robust fault detection},
  journal = {Automatica},
  volume    = {39},
  number    = {6},
  pages     = {1095--1102},
  year      = {2003},
}

@article{seeber2024implicit,
  author       = {Seeber, Richard},
  title        = {Implicit discrete‑time implementation of robust exact differentiators--a toolbox},
  journal = {at-Automatisierungstechnik},
  year         = {2024},
  volume       = {72},
  number       = {8},
  pages        = {757--768},
  doi          = {10.1515/auto‑2024‑5068},
}

@article{aldana2025optimal,
  author       = {Aldana-L{\'o}pez, Rodrigo and Seeber, Richard and Haimovich, Hernan and G{\'o}mez-Guti{\'e}rrez, David},
  title        = {Optimal robust exact first-order differentiators with {L}ipschitz-continuous output},
  journal = {IEEE Trans. Autom. Control},
  year         = {2025},
  volume       = {70},         
  number       = {6},          
}

@inproceedings{diop2000numerical,
  author       = {S. Diop and J.~W. Grizzle and F. Chaplais},
  title        = {On numerical differentiation algorithms for nonlinear estimation},
  booktitle = {Proc. 39th IEEE CDC},
  year         = {2000},
  pages        = {1133--1138}
}

@article{belanger1998velocity,
  author       = {P.~R. Bélanger and P. Dobrovolny and A. Helmy and X. Zhang},
  title        = {Estimation of Angular Velocity and Acceleration from Shaft‑Encoder Measurements},
  journal = {Int. J. Robot. Res.},
  year         = {1998},
  volume       = {17},
  number       = {11},
  pages        = {1225--1233},
}

@article{fridman2011,
  author    = {Emmanuel Cruz-Zavala and Jaime A Moreno and Leonid M Fridman},
  title     = {Uniform robust exact differentiator},
  journal = {IEEE Trans. Autom. Control},
  volume    = {56},
  number    = {11},
  pages     = {2727--2733},
  year      = {2011},
}

@book{Luenberger,
  title={Optimization by Vector Space Methods},
  author={Luenberger, David G.},
  year={1969},
  publisher={John Wiley \&{} Sons}
}

@book{EkelandTemam,
  title={Convex Analysis and Variational Problems},
  author={Ekeland, Ivar and Temam, Roger},
  year={1999},
  publisher={SIAM}
}

@book{Brezis,
  title={Functional Analysis, Sobolev Spaces and Partial Differential Equations},
  author={Br{\'e}zis, Haim},
  year={2011},
  publisher={Springer}
}

@article{Vasiljevic2008,
  author  = {Luma K. Vasiljevic and Hassan K. Khalil},
  title   = {Error bounds in differentiation of noisy signals by high-gain observers},
  journal = {Syst. Control Lett.},
  volume  = {57},
  number  = {10},
  pages   = {856--862},
  year    = {2008}
}

@incollection{Moreno2023,
  author    = {Jaime A. Moreno},
  title     = {Bi-homogeneous Differentiators},
  booktitle = {Sliding-Mode Contr. and Variable-Structure Syst.: The State of the Art},
  editor    = {Tiago Roux Oliveira and Leonid Fridman and Liu Hsu},
  series    = {Studies in Systems, Decision and Control},
  volume    = {490},
  pages     = {71--96},
  publisher = {},
  address   = {Cham},
  year      = {2023}
}

@article{SavGol,
  author    = {Michael Schmid and David Rath and Ulrike Diebold},
  title     = {Why and How {S}avitzky--{G}olay Filters Should Be Replaced},
  journal = {ACS Meas. Sci. Au},
  volume    = {2},
  number    = {2},
  pages     = {185--196},
  year      = {2022}
}

@article{nemirovski1984signal,
  title={Signal processing by the nonparametric maximum-likelihood method},
  author={Nemirovski, Arkadi S. and Polyak, Boris T. and Tsybakov, Aleksandr B.},
  journal = {Probl. Inf. Transm.},
  volume={20},
  number={3},
  pages={29--46},
  year={1984},
  publisher={Russian Academy of Sciences}
}

@misc{MLEDiff,
  author = {Konstantin Avrachenkov and Leonid Freidovich},
  title = {Maximum-likelihood-estimator-based differentiator},
  howpublished = "\url{https://github.com/fleonid/MLEdiff}",
  year = {2025}
}

@article{Livne,
  title={Proper discretization of homogeneous differentiators},
  author={Livne, Mike and Levant, Arie},
  journal = {Automatica},
  volume={50},
  pages={2007--2014},
  year={2014},
  publisher={Elsevier}
}

@article{Kolmogorov1962,
  author    = {Andrey N. Kolmogorov},
  title     = {On Inequalities Between Upper Bounds of Consecutive Derivatives of an Arbitrary Function Defined on an Infinite Interval},
  journal = {Amer. Math. Soc. Transl.},
  series    = {Series 2},
  volume    = {9},
  pages     = {233--242},
  year      = {1962}
}

@article{dabroom2001,
  title={Discrete-time implementation of high-gain observers for numerical differentiation},
  author={Dabroom, Ahmed M. and Khalil, Hassan K.},
  journal = {Int. J. Control},
  volume={72},
	number={17},
  pages={1523--1537},
  year={1999},
  publisher={Taylor & Francis Group}
}

@article{Barbot2020discrete,
  title={Discrete differentiators based on sliding modes},
  author={Barbot, Jean-Pierre and Levant, Arie and Livne, Miki and Lunz, Davin},
  journal = {Automatica},
  volume={112},
  pages={108633},
  year={2020},
  publisher={Elsevier}
}

@phdthesis{Carvajal2022contribution,
  title={Contribution to the Discretization of Sliding Mode Differentiators},
  author={Carvajal Rubio, Jos{\'e} Eduardo},
  year={2022},
  school={Universit{\'e} Polytechnique Hauts-de-France}
}

@article{Carvajal2022implicit,
  author    = {Jos{\'e} Eduardo Carvajal-Rubio and Michael Defoort and Juan Diego S{\'a}nchez-Torres and Mohamed Djemai and Alexander G. Loukianov},
  title     = {Implicit and explicit discrete-time realizations of the robust exact filtering differentiator},
  journal = {J. Franklin Inst.},
  volume    = {359},
  number    = {8},
  pages     = {3951--3978},
  year      = {2022},
  doi       = {10.1016/j.jfranklin.2022.01.015},
  publisher={Elsevier}
}

@article{Chartrand2011numerical,
  author  = {Rick Chartrand},
  title   = {Numerical Differentiation of Noisy, Nonsmooth Data},
  journal = {ISRN Appl. Math.},
  pages   = {Article ID 164564},
  year    = {2011}
}

@article{Cullum1971numerical,
  title={Numerical differentiation and regularization},
  author={Cullum, Jane},
  journal = {SIAM J. Numer. Anal.},
  volume={8},
  number={2},
  pages={254--265},
  year={1971},
  publisher={SIAM}
}

@article{Levant1998robust,
  title={Robust exact differentiation via sliding mode technique},
  author={Levant, Arie},
  journal = {Automatica},
  volume={34},
  number={3},
  pages={379--384},
  year={1998},
  publisher={Elsevier}
}

@article{Mboup2009numerical,
  title={Numerical differentiation with annihilators in noisy environment},
  author={Mboup, Mamadou and Join, C{\'e}dric and Fliess, Michel},
  journal = {Numer. Algorithms},
  volume={50},
  pages={439--467},
  year={2009},
  publisher={Springer}
}

@article{Mboup2018frequency,
  title={Frequency-domain analysis and tuning of the algebraic differentiators},
  author={Mboup, Mamadou and Riachy, Samer},
  journal = {Int. J. Control},
  volume={91},
  number={9},
  pages={2073--2081},
  year={2018},
  publisher={Taylor and Francis}
}

@article{Othmane2022survey,
  title={Survey on algebraic numerical differentiation: historical developments, parametrization, examples, and applications},
  author={Othmane, Amine and Kiltz, Lothar and Rudolph, Joachim},
  journal = {Int. J. Syst. Sci.},
  volume={53},
  number={9},
  pages={1848--1887},
  year={2022},
  publisher={Taylor and Francis}
}

@article{Seeber2023,
title = {Optimal robust exact differentiation via linear adaptive techniques},
journal = {Automatica},
volume = {148},
pages = {110725},
year = {2023},
author = {Richard Seeber and Hernan Haimovich},
}
\bibliographystyle{IEEEtran}

\end{document}